\documentclass[english,aps,notitlepage]{revtex4-1}
\usepackage{ae,aecompl}
\usepackage[T1]{fontenc}
\usepackage[latin9]{inputenc}
\usepackage{babel}
\usepackage{units}
\usepackage{amsmath}
\usepackage{amssymb}
\usepackage{graphicx}
\usepackage{esint}
\usepackage[T1]{fontenc}
\usepackage[latin9]{inputenc}
\usepackage{stmaryrd}
\usepackage{babel}
\usepackage[colorlinks,citecolor=cyan,linkcolor=blue,urlcolor=blue]{hyperref}
\usepackage{babel}

\makeatother

\begin{document}
\preprint{APS/123-QED}

\title{Black-Body Radiation in an Accelerated Frame}
\author{Seramika Ariwahjoedi$^{1,2}$}
\email{seramika.wahyoedi@apctp.org}

\author{Apriadi Salim Adam$^{2}$}
\email{apriadi.salim.adam@brin.go.id }

\author{Hadyan Luthfan Prihadi$^{3,4}$}
\email{hadyanluthfan@itb.ac.id }

\author{Freddy Permana Zen$^{3,4}$}
\email{fpzen@fi.itb.ac.id }

\address{$^{1}$Asia Pacific Center for Theoretical Physics,\\ Pohang University
of Science and Technology, Pohang 37673, Gyeongsangbuk-do, South Korea,}

\address{$^{2}$Research Center for Quantum Physics,\\ National Research and
Innovation Agency (BRIN), South Tangerang 15314, Banten, Indonesia,
}

\address{$^{3}$Theoretical High Energy Physics Group, Department of Physics,\\ FMIPA, Institut Teknologi
Bandung, Jl. Ganesha 10 Bandung 40132, West Java, Indonesia,}

\address{$^{4}$Indonesia Center for Theoretical and Mathematical Physics
(ICTMP), Indonesia.~\\}

\date{\today}

\begin{abstract}
We derive Planck's radiation law in a uniformly accelerated frame
expressed in Rindler coordinates. The black-body spectrum is time-dependent
by its temperature and Planckian at each instantaneous time, but it
is scaled by an emissivity factor that depends on the Rindler spatial
coordinate and the acceleration magnitude. The observer in an accelerated
frame will perceive the black-body as black, hyperblack, or grey,
depending on its position with respect to the source (moving away
or towards), the acceleration magnitude, and the case of whether it
is accelerated or decelerated. For an observer accelerating away from
the source, there exists a threshold on the acceleration magnitude
beyond which it stops receiving radiation from the black-body. Since
the frequency and the number of modes in Planck's law evolve over
time, the spectrum is continuously red or blue-shifted towards lower
(or higher) frequencies as time progresses, and the radiation modes
(photons) could  be created or annihilated, depending on the observer's
position and its acceleration or deceleration relative to the source
of radiation.
\end{abstract}
\maketitle

\section{Introduction}

The status of temperature in relativistic thermodynamics remained
a subject of ongoing debate, marked by the lack of a universal agreement
in the physics community regarding the transformation of temperature
for moving bodies \citep{Farias,Derakshani}. In their works, both
Planck and Einstein independently derived that the temperature of
a moving body would transform as $T'=\frac{T}{\gamma}$, with $\gamma$
representing the Lorentz factor associated with the velocity of the
moving body (or moving observer, if the body is at rest) \citep{Planck,Einstein,Einstein2}.
However, under different assumptions, Ott and Arzeli$\mathrm{\grave{e}}$s
derived the transformation of temperature to be $T'=\gamma T$ \citep{Ott,Arzelies}.
Furthermore, Landsberg proposed the invariance of temperature under
transformations between inertial frames, namely, $T=T'$ \citep{Landsberg,Lansberg2}.
Each of these different formulation for temperature transformations
in inertial frames is supported by compelling arguments, hence, although
the most widely-accepted theory is the one by Einstein-Planck \citep{Einstein,Planck},
the attempt to reconcile these different approaches remains an ongoing
research. A recent advancement in the field of special relativistic
thermodynamics was achieved by Nakamura in \citep{Nakamura,Nakamura2}
who revised the Israel-van Kampfen covariant, inverse, 4-temperature
\citep{vanKampen,Israel}. This revision allows the derivation of
the three distinct temperature transformations within the Israel-van
Kampen formulation, depending on the definition of the 3-dimensional
volume and the chosen decomposition of the 4-momentum \citep{Nakamura,Nakamura2}.

Moreover, Tolman pioneered the study of general relativistic thermodynamics
in \citep{Tolman,Tolman2}, by proposing a law that in a gravitational
field, the temperature of a system in thermal equilibrium is inversely
proportional to the square root of the gravitational potential at
that location. This law can be expressed as $T'=T\sqrt{g_{tt}}$,
where $g_{tt}$ represents the purely-time component of the metric
tensor $g_{\mu\nu}$. In general relativity, the metric tensor describes
the gravitational field as the curvature of spacetime, and the $g_{tt}$
component corresponds to the gravitational potential in the Newtonian
limit \citep{Einstein-1,Carrol}. 

On the other hand, the concept of temperature can be defined from
several theoretical framework. In thermodynamics, temperature is defined
as the parameter shared by two bodies in thermal equilibrium, a state
characterized by the absence of net heat transfer between the bodies
\citep{Salinger}; this definition is widely accepted
among $19^{\mathrm{th}}$-century physicists including Clausius, Carnot,
Kelvin, and many others. From the statistical mechanics approach,
it is defined as the average of kinetic energy of a many-body system,
and this is described by the relation $\left\langle E\right\rangle \sim\kappa_{B}T$,
with $\kappa_{B}$ is the Boltzmann constant \citep{Boltzmann}. Statistical
mechanics provides a procedure for bridging macroscopic observables
of a system, such as temperature and entropy, to its microscopic properties,
i.e., the position, velocity, and kinetic energy of the individual
degrees of freedom \citep{Liboff}. Combining these two perspectives,
temperature can be defined as an observable that emerge from the statistical
properties of individual degrees of freedom (i.e., the kinetic energy)
of a many-body system that is in a thermal equilibrium with its surrounding.
As far as we are concern, this is the current status of the temperature
conceptualization widely-accepted by physics community at the moment.

In our study, we adopt a more pragmatic approach to the concept of
temperature -specifically, through the use of Planck's law on black-body
radiation, originally proposed in 1900 \citep{Planck2}. This law
initiated the birth of quantum mechanics and remains valid through
experiments to the present day, and it founds wide applications in
pyrometric measurements. As explained in the preceding paragraphs,
the thermodynamic definition of temperature -a parameter indicating
two systems are in thermodynamic equilibrium- is particularly applicable
for temperature measurement in a rest frame. The measurement apparatus
must establish direct contact with the system and afford sufficient
time for thermalization until a (thermal) equilibrium is reached.
In this case, the temperature is well-defined. However, for these
two systems to be in thermal equilibrium, in prior, they need to be
in a mechanical equilibrium, while maintaining a direct contact to
undergo thermalization. This requirement is challenging to fulfill
if one of the systems is not at rest relative to the other \citep{vanKampen,vanKampen3}.
In principle, at least by the definition originating from thermodynamical
approach, the notion of temperature can only emerge as a consequence
of local measurement \citep{Mares}. Furthermore, a quantum-field
theory calculation, using the Unruh-DeWitt detector as a thermometer
moving through a thermal bath, gives a result that the particle distribution
is non-Planckian, which made it difficult to define temperature in
this context \citep{Costa}. Based on this work, then Landsberg and
Matsas proposed an argument on the impossibility of a universal relativistic
temperature transformation, due to the fact that there is no continuous
function that could map the non-Planckian distribution from \citep{Costa}
to a Planckian one \citep{Matsas,Matsas2}. These reasons supports
the argument that the concept of temperature of a moving body is not
well-defined \citep{Mares}.

However, using the Planckian spectrum, one can predict the temperature
of a distant object without necessitating direct contact, thereby
skipping the need to reach thermal equilibrium with the object under
observation. This methodology, treating temperature as a derived parameter,
finds widespread application \textendash{} from the infrared thermometers
to the calculation of the temperatures of celestial bodies such as
stars and the cosmic microwave background \citep{Bbracewell}. The
process involves the collection of the complete radiation spectrum
emitted by the object and, -under the assumption that the object behaves
as a black-body-, the comparation of the spectrum data with the theoretical
black-body spectrum. Temperature prediction can be carried out using
either Wien's displacement law \citep{Wien,Wien2} or the Stefan-Boltzmann
law \citep{Stefan}. In the case of a black-body source, these two
distinct temperature calculation methods coincide. However, it is
important to note that assuming all objects behave as perfect black-bodies
is overly restrictive. For a more realistic approach, the collected
data should be compared to a calibrated grey-body spectrum derived
from the black-body spectrum, but with emissivity or absorptivity
values set below 1. This method, to the best of our knowledge, represents
the sole viable approach for 'measuring' the temperature of a moving
object. Furthermore, experiments based on this measurement technique
are relatively straightforward to conduct. These facts emphasize the
importance of Planck's law in the determination of temperatures of
moving bodies.

We derived Planck's law and calculated the black-body spectrum in
a uniformly accelerated frame expressed in Rindler coordinates. The
reason behind our quest on this subject can be explained as follows:
In the framework of general relativity, an inertial frame is an idealization,
hence one needs to move to a more general reference of frame. As a
first realistic step towards achieving this objective, we consider
a uniformly accelerated frame, expressed in Rindler coordinates. The
calculation of the black-body spectrum in the inertial frame had been
done in \citep{Peebles,Heer,Henry}. These works demonstrated that
the Planckian spectrum is invariant under Lorentz transformation,
and the effective temperature of the moving object depends on its
velocity and the direction of observation \citep{Peebles,Heer,Henry}.
Another different approach shows only the zero-point temperature term
is invariant under Lorentz transformation \citep{Bouyer}. Attempts
to calculate the black-body spectrum in an accelerated frame had been
done in \citep{Lee}, but to the best of our knowledge, the calculation
of the spectrum in Rindler coordinates is still lacking in the existing
literature. In the present work, we follow the (effective) 'directional-temperature'
approach as in \citep{Henry}, but instead of using the temperature
$T$ in Planck's law, we substitute it with variables that maximize
the energy in the spectrum (in our case, we use the 'maximal' frequency).
With this substitution, we avoid the problem associated with temperature
transformation, as we are certain about the Lorentz transformation
for the 'maximal' variables.

The main result presented in this article is the black-body spectrum
in an accelerated frame. The spectrum has an explicit time dependency
on its temperature, Planckian at each instantaneous time, but it is
scaled by a factor dependent on the spatial coordinate and the acceleration
magnitude. The spatial, coordinate-dependent scale factor is proportional
to $\sim e^{\alpha\xi}$, while the scale factor related to the acceleration
is proportional to $\sim e^{\alpha\xi}\left(1\pm\alpha\right)$, where
the $\pm$ sign depending on the observer is either deccelerated or
accelerated. This scale factor could be physically interpreted as
the emissivity factor of the source, hence, in an accelerated frame,
a black-body could be perceived as grey or 'hyper-black', depending
on the magnitude of the acceleration. Furthermore, the variables in
Planck's law, specifically the number of modes and frequency, evolve
over time. However, since we are considering all the possible value
of frequencies from zero to infinity, the time dependence of the frequencies
does not explicitly shown in the law. The Planckian spectrum is continuously
red or blue-shifted towards lower (or higher) frequencies as time
progresses. In the accelerated frame, we demonstrate that the radiation
modes (photons) could be positive or negative, depending on the acceleration
or deceleration of the observer, and zero for vanishing acceleration.
In the end, assuming the validity of Wien's displacement law in an
accelerated frame, the time-dependent, (directional)-temperature of
a body in an accelerated frame is given by $\bar{T}\left[\tau\right]=Te^{-\alpha\xi}\left(\cosh\bar{\varsigma}\left[\tau\right]+\cos\bar{\theta}\left[\tau\right]\sinh\bar{\varsigma}\left[\tau\right]\right)^{-1}$.
It is worth noting that the spatial, coordinate-dependent scale factor
$\sim e^{-\alpha\xi}$ resembles the temperature scale factor in Tolman's
theory of general relativistic thermodynamics \citep{Tolman}, and
for $\alpha=0,$ the transformation of the temperature returns to
the ones obtained in \citep{Peebles,Heer,Henry}.

The paper is structured as follows: Section II contains the derivation
of black-body radiation in an accelerated frame. First, we introduce
the properties of the Rindler space, followed by the transformation
from Minkowski to Rindler coordinates for each variable contained
in Planck's law. With these transformed variables, we derive Planck's
distribution law in the accelerated frame. Section III contains discussions
and the physical interpretation of the results, including the aberration
of light, relativistic beaming, and red/blueshifts due to the relativistic
Doppler effect, creation/annihilation of modes in the accelerated/decelerated
frame, the emissivity factor of the black-body source, and the transformation
of temperature, assuming Wien's law holds in the accelerated frame.
Finally, in Section IV, we provide a conclusive summary of our work
and outline some insights into potential further research. We have
included appendices to ensure the self-contained nature of this paper.
The appendices contain the derivation of the black-body spectrum in
a frame moving with a constant velocity. This supplementary content
aids readers in understanding each step of the spectrum derivation
in the accelerated frame. One crucial aspect is that we have aimed
to keep this paper as classical as possible. The only quantum assumptions
we adopt are the explicit use of the de Broglie postulate $\,^{3}\mathrm{p}=\hbar\mathrm{k}$
for photon momentum (with $\mathrm{k}$ is the photon wave-vector
and $\hbar$ is the Planck constant) and the assumption of energy
discreteness $E=n\hbar\omega$ (with $\omega$ is the photon angular
frequency and $n$ are integers) as the requirement to derive Planck's
law.

\section{Blackbody Radiation in Accelerated Frame}

\subsection{Rindler Coordinates}

Let $\left(\mathcal{M},\mathbf{x}\right)$ be a Minkowski space equipped
with a metric $\eta$. The (global) coordinate $\mathbf{x}$ =$\left(x,y,z,t\right)$,
with $t:=c\mathfrak{t}$ is the time coordinate and $\left(x,y,z\right)$
is the spatial part, is used to parametrize $\mathcal{M}$. In this
coordinate, the metric $\eta$ is a diagonal metric with signature
$\left(-1,1,1,1\right)$. Naturally, one could attach an inertial
reference of frame to the coordinate $\mathbf{x}$, and let $\mathcal{O}$
be an observer at rest with respect to this inertial frame. Let us
call this frame/observer as $\mathcal{O}.$

Suppose we have another observer $\mathcal{A}$ that moves with a
constant acceleration of magnitude $a$ in the direction of the $x$-axis,
with respect to $\mathcal{O}$. The trajectory of $\mathcal{A}$ in
$\mathcal{M}$ is $\mathbf{x}\left[\lambda\right]=x^{\mu}\left[\lambda\right]$,
with $\lambda$ is a parameter (a proper time with respect to $\mathcal{A}$),
satisfying:
\begin{align}
t\left[\lambda\right] & =\frac{1}{a}\sinh a\lambda,\label{eq:from1}\\
x\left[\lambda\right] & =\frac{1}{a}\cosh a\lambda,\label{eq:from2}
\end{align}
and $y,z$ being constant. The 4-acceleration is $a^{\mu}=\frac{d^{2}x^{\mu}}{d\lambda^{2}},$
where the non-zero components are:
\begin{align*}
a^{t} & =a\sinh a\lambda,\\
a^{x} & =a\cosh a\lambda,
\end{align*}
such that the magnitude of $a^{\mu}$ is $\sqrt{a^{\mu}a_{\mu}}=a$.
Note  here that $a$ is positive definite: $a\geq0$. From (\ref{eq:from1})
and (\ref{eq:from2}), the trajectory of $\mathcal{A}$ satisfies
$x^{2}-t^{2}=a^{-2}$, describing a collection of hyperboloids with
asymptotes at the lines $x=-t$ and $x=t$. The asymptotes divide
$\mathcal{M}$ into 4 regions, and let us focus on one of the region
where $0<x<\infty$ and $-x<t<x$, usually labeled as the (right)
Rindler wedge, see FIG. 1. 
\begin{figure}
\begin{centering}
\includegraphics[scale=0.65]{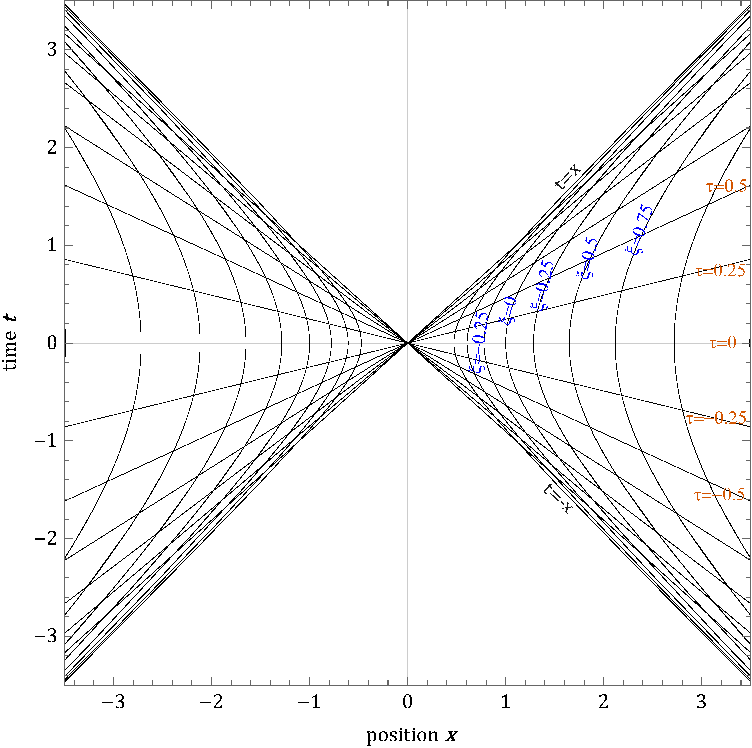}
\par\end{centering}
\caption{The left and right Rindler wedge. The hyperboloids describe trajectories
of constant $\xi$, while the straight lines of varying slopes describe
trajectories of constant $\tau$. The point $\left(x,t\right)=\left(0,0\right)$
in the Cartesian coordinates coincides with the points $\left(\tau,\xi\right)=\left(0,-\infty\right)$
in the Rindler coordinate. The (Killing) horizons are located at $x=\pm t$.}
\end{figure}
 One could choose a new coordinate patch $\bar{\mathbf{x}}$ =$\left(\xi,\bar{y},\bar{z},\tau\right)$,
to parametrize the Rindler wedge as follows:
\begin{equation}
\begin{array}{cc}
\tau= & \frac{1}{2\alpha}\left(\ln\alpha\left(x+t\right)-\ln\alpha\left(x-t\right)-2\varsigma\right),\\
\xi= & \frac{1}{2\alpha}\left(\ln\alpha\left(x+t\right)+\ln\alpha\left(x-t\right)\right),
\end{array}\label{eq:Rindler}
\end{equation}
and their inverse:
\begin{equation}
\begin{array}{cc}
t & =\frac{1}{\alpha}e^{\alpha\xi}\sinh\left(\alpha\tau+\varsigma\right),\\
x & =\frac{1}{\alpha}e^{\alpha\xi}\cosh\left(\alpha\tau+\varsigma\right),
\end{array}\label{eq:invRindler}
\end{equation}
with $\bar{y}=y$ , $\bar{z}=z$, and $\alpha,$ $\varsigma$ are
constant parameters. The (right) Rindler wedge is covered by the coordinate
patch with $-\infty<\xi,\tau<\infty$. For $\varsigma=0$, the coordinate
is usually known as the Rindler coordinates  in the Lass/radar representation
\citep{Carrol}. In this coordinate, (\ref{eq:from1})-(\ref{eq:from2})
becomes:
\begin{align*}
\tau\left[\lambda\right] & =\frac{1}{\alpha}\left(a\lambda-\varsigma\right),\\
\xi\left[\lambda\right] & =\frac{1}{\alpha}\ln\frac{\alpha}{a}.
\end{align*}
Notice that along the trajectory $\mathbf{x}\left[\lambda\right]$,
$\xi$ is constant and $\tau$ is proportional to the proper time
$\lambda$, although it is shifted by a constant amount $-\frac{\varsigma}{\alpha}$.
An observer moving along constant $\xi$ will experience an acceleration
of magnitude:
\begin{equation}
a=\alpha e^{-\alpha\xi}.\label{eq:magnitude}
\end{equation}
Since $a$ is positive definite, then so does $\alpha$: $\alpha\geq0$.
 Naturally, the coordinate $\bar{\mathbf{x}}$ is adopted by the observer/frame
$\mathcal{A}$.

One can  obtain the infinitesimal transformation of (\ref{eq:Rindler}),
which is:
\begin{align}
d\tau & =\frac{xdt-tdx}{\alpha\left(x^{2}-t^{2}\right)}=e^{-\alpha\xi}\left(\cosh\bar{\varsigma}dt-\sinh\bar{\varsigma}dx\right),\label{eq:Rindler-1}\\
d\xi & =\frac{xdx-tdt}{\alpha\left(x^{2}-t^{2}\right)}=e^{-\alpha\xi}\left(\cosh\bar{\varsigma}dx-\sinh\bar{\varsigma}dt\right),\nonumber 
\end{align}
where we write $\bar{\varsigma}=\bar{\varsigma}\left[\tau\right]=\alpha\tau+\varsigma$,
and their inverses:
\begin{align}
dt & =e^{\alpha\xi}\left(\cosh\bar{\varsigma}d\tau+\sinh\bar{\varsigma}d\xi\right)=\alpha\left(xd\tau+td\xi\right),\nonumber \\
dx & =e^{\alpha\xi}\left(\cosh\bar{\varsigma}d\xi+\sinh\bar{\varsigma}d\tau\right)=\alpha\left(xd\xi+td\tau\right),\label{eq:Rindler-2}
\end{align}
with $d\bar{y}=dy$ and $d\bar{z}=dz$. $\bar{\varsigma}$ is the
time-dependent rapidity that is linear to the ``proper time'' $\tau$.
For $\alpha=0$, the acceleration $a$ vanishes, and (\ref{eq:Rindler-1})-(\ref{eq:Rindler-2})
will return to the standard Lorentz transformation (\ref{eq:Lorentz-2})-(\ref{eq:Lorentz-3}).

The line element in Minkowksi space could be written in Rindler coordinate
using (\ref{eq:Rindler-2}) as follows:
\begin{equation}
ds^{2}=g_{\mu\nu}d\bar{x}^{\mu}d\bar{x}^{\nu}=e^{2\alpha\xi}\left(-d\tau^{2}+d\xi^{2}\right)+d\bar{y}^{2}+d\bar{z}^{2};\label{eq:Rindlermetric}
\end{equation}
The coefficient of the line element gives the Rindler metric $g_{\mu\nu}$
in Lass representation. One could also consider the left Rindler wedge
(the region IV in FIG. 1), where it could be covered by a coordinate
patch similar to the RRW, but with the flip in the sign of (\ref{eq:invRindler})
as follows:
\begin{equation}
\begin{array}{cc}
t & =-\frac{1}{\alpha}e^{\alpha\xi}\sinh\left(\alpha\tau+\varsigma\right),\\
x & =-\frac{1}{\alpha}e^{\alpha\xi}\cosh\left(\alpha\tau+\varsigma\right).
\end{array}\label{eq:invRindlerLRW}
\end{equation}
In this region, the future-directed time-like Killing vector is $-\partial\tau$,
instead of $\partial\tau$ in region I.

\subsection{The Aberration of Light in Accelerated Frame}

Suppose with respect to an inertial frame $\mathcal{O}$ we have an
object moving with a constant velocity $\mathbf{v}$ as follows:
\[
\mathbf{v}=\left(v_{x},v_{y},v_{z}\right)=\left(\frac{dx}{dt},\frac{dy}{dt},\frac{dz}{dt}\right).
\]
According to the accelerated frame $\mathcal{A}$, the velocity of
the object is:
\[
\bar{\mathbf{v}}\left[\tau,\xi\right]=\left(v_{\xi},\bar{v}_{y},\bar{v}_{z}\right)=\left(\frac{d\xi}{d\tau},\frac{d\bar{y}}{d\tau},\frac{d\bar{z}}{d\tau}\right).
\]
Notice that the components of $\bar{\mathbf{v}}\left[\tau,\xi\right]$
are functions of the Rindler coordinate $\left(\tau,\xi\right)$.
Using the transformation (\ref{eq:Rindler-1})-(\ref{eq:Rindler-2}),
one can obtain the relation between the instantaneous $\bar{\mathbf{v}}\left[\tau,\xi\right]$
and $\mathbf{v}$ as follows:
\begin{equation}
v_{\xi}=\frac{v_{x}-\tanh\bar{\varsigma}}{1-v_{x}\tanh\bar{\varsigma}}=\frac{\left(xv_{x}-t\right)}{\left(x-tv_{x}\right)},\label{eq:velocity3}
\end{equation}
\begin{equation}
\bar{v}_{y,z}=\frac{e^{\alpha\xi}v_{y,z}}{\cosh\bar{\varsigma}\left(1-v_{x}\tanh\bar{\varsigma}\right)}=\frac{\alpha\left(x^{2}-t^{2}\right)v_{y,z}}{\left(x-tv_{x}\right)}.\label{eq:velocity4}
\end{equation}
Relation (\ref{eq:velocity3})-(\ref{eq:velocity4}) could be derived
from the standard velocity addition formula (or, similarly, from the
relativistic acceleration formula). Their inverses could also be obtained
as follows:
\[
v_{x}=\frac{v_{\xi}+\tanh\bar{\varsigma}}{1+v_{\xi}\tanh\bar{\varsigma}},
\]
\[
v_{y,z}=\frac{\bar{v}_{y,z}}{e^{\alpha\xi}\cosh\bar{\varsigma}\left(1+v_{\xi}\tanh\bar{\varsigma}\right)}.
\]

In the following paragraphs, we will derive the light aberration formula
for an accelerated observer $\mathcal{A}$. For a brief review on
the light aberration formula for an inertial, moving observer, one
could refer to  Appendix B, or consult \citep{Johnson}. Let us consider
a black-body source (photons inside a cavity) at rest with respect
to an inertial frame $\mathcal{O}.$ Electromagnetic radiations (photons)
are emitted by the black-body with propagation velocity $c$ in the
direction of $\hat{n}=\left(\cos\theta,\sin\theta,0\right)$, with
$\theta$ is the (polar) observation (or inclination) angle between
the axis $x$ and $\hat{n}$ on plane $dx\wedge dy$. The velocity
of the photon with respect to frame $\mathcal{O}$ is:
\begin{equation}
\mathbf{v}=\left(v_{x},v_{y},v_{z}\right)=c\hat{n}=\left(c\cos\theta,c\sin\theta,0\right).\label{eq:velocity5}
\end{equation}
The 4-velocity of the photon satisfies the null-vector condition,
see Appendix X. To derive the 3-velocity in $\mathcal{A}$, let us
first write the 4-velocity in Rindler coordinate:
\[
\mathbf{u}=\bar{\mathbf{u}}=\frac{\partial\tau}{\partial\lambda}\left(1,\bar{\mathbf{v}}\right),
\]
where $\lambda$ is a real parameter (usually the proper time with
respect to the moving object) and $\bar{\mathbf{v}}$ is the 3-velocity
in $\mathcal{A}$:
\[
\bar{\mathbf{v}}=\left(v_{\xi},\bar{v}_{y},\bar{v}_{z}\right)=\left(\frac{\partial\xi}{\partial\tau},\frac{\partial\bar{y}}{\partial\tau},\frac{\partial\bar{z}}{\partial\tau}\right).
\]
Using the null-vector condition for light, we have:
\[
\bar{u}^{\mu}\bar{u}_{\mu}=g_{\mu\nu}\bar{u}^{\mu}\bar{u}^{\nu}=\left(\frac{\partial\tau}{\partial\lambda}\right)^{2}\left(-e^{2\alpha\xi}+e^{2\alpha\xi}v_{\xi}^{2}+\bar{v}_{y}^{2}+\bar{v}_{z}^{2}\right)=0.
\]
In $\mathcal{O}$, the light only propagates in the direction $\hat{n}$
on plane $dx\wedge dy$, so even in $\mathcal{A}$, the $\bar{z}$
component of the velocity of light is zero. The null condition then
becomes:
\[
v_{\xi}^{2}+\left(e^{-\alpha\xi}\bar{v}_{y}\right)^{2}=1.
\]
Writing $v_{\xi}^{2}=\cos^{2}\bar{\theta}$ and $\left(e^{-\alpha\xi}\bar{v}_{y}\right)^{2}=\sin^{2}\bar{\theta},$
the 3-velocity of the photon in $\mathcal{A}$ is:
\begin{equation}
\bar{\mathbf{v}}=\left(v_{\xi},\bar{v}_{y},\bar{v}_{z}\right)=c\hat{\bar{n}}=\left(c\cos\bar{\theta},ce^{\alpha\xi}\sin\bar{\theta},0\right),\label{eq:velocity6}
\end{equation}
with $\hat{\bar{n}}$ and $\bar{\theta}$ are the propagation direction
and the inclination angle of the photon at $\mathcal{A}$, respectively.
Notice that since the observer $\mathcal{A}$ is accelerated in the
direction $+x$, it will receive only photons that are moving towards
the observer, namely, the ones that have the $+x$ component in the
velocity. The relation between $\bar{\mathbf{v}}$ and $\mathbf{v}$
can be obtained using the velocity addition formula in the accelerated
frame (\ref{eq:velocity3})-(\ref{eq:velocity4}). Inserting (\ref{eq:velocity5})-(\ref{eq:velocity6})
to (\ref{eq:velocity3})-(\ref{eq:velocity4}) will result in the
light aberration for an accelerated frame $\mathcal{A}$:
\begin{align}
\cos\bar{\theta} & =\frac{\cos\theta-\tanh\bar{\varsigma}}{1-\cos\theta\tanh\bar{\varsigma}},\label{eq:aberrationrindler}\\
\sin\bar{\theta} & =\frac{\sin\theta}{\cosh\bar{\varsigma}\left(1-\cos\theta\tanh\bar{\varsigma}\right)},\label{eq:ab}\\
\tan\frac{\bar{\theta}}{2} & =\frac{1}{\cosh\bar{\varsigma}\left(1-\tanh\bar{\varsigma}\right)}\tan\frac{\theta}{2},
\end{align}
where the last equation is obtained from the trigonometry identity
$\tan\frac{\theta}{2}=\frac{\sin\theta}{1+\cos\theta}$. One could
also obtained their inverses:
\begin{align}
\cos\theta & =\frac{\cos\bar{\theta}+\tanh\bar{\varsigma}}{1+\cos\bar{\theta}\tanh\bar{\varsigma}},\label{eq:vabe2}\\
\sin\theta & =\frac{\sin\bar{\theta}}{\cosh\bar{\varsigma}\left(1+\cos\bar{\theta}\tanh\bar{\varsigma}\right)},\label{eq:aberrationrindler2}\\
\tan\frac{\theta}{2} & =\frac{1}{\cosh\bar{\varsigma}\left(1+\tanh\bar{\varsigma}\right)}\tan\frac{\bar{\theta}}{2}.
\end{align}
The aberration formula in $\mathcal{A}$ has exactly the same form
with the ones in a constant moving frame $\mathcal{O}'$ (see relation
(\ref{eq:aberration-1})-(\ref{eq:aberration-3})), however, it should
be kept in mind that $\bar{\theta}$ is time-dependent, namely $\bar{\theta}=\bar{\theta}\left[\tau\right]$,
while $\theta'$  is constant.

\subsection{The Doppler Effect in Accelerated Frame}

In this paper we will only consider the longitudinal relativistic
Doppler effect \citep{LongitudDoppler}, where the observer/source
velocity has component parallel to the wave propagation. To derive
this, let us first consider the 4-momentum constraint (\ref{eq:energyconstraint})
which is valid for any reference frame. The photon is massless, so
using the de Broglie postulate $E=\hbar\omega$ and $\mathbf{p}=\hbar\mathbf{k}$,
(\ref{eq:energyconstraint}) becomes the dispersion relation:
\begin{equation}
\omega^{2}=\left|\mathbf{k}\right|^{2},\label{eq:disperssionrel}
\end{equation}
for the wave vector of the photon satisfies $\mathbf{k}=\frac{2\pi\hat{n}}{\lambda}.$
In $\mathcal{O},$ the dispersion relation becomes:
\[
\omega^{2}=k_{x}^{2}+k_{y}^{2}.
\]
Since the coordinate axis $x$ is perpendicular to $y$, one could
write $k_{x}=\omega\cos\theta$ and $k_{y}=\omega\sin\theta$, with
$\theta$ is the inclination angle at $\mathcal{O}.$ Therefore, the
 4-momentum of a photon with frequency $\omega$ according to the
rest frame $\mathcal{O}$ (with respect to the black-body source)
is $\mathbf{p}=\left(p_{t},p_{x},p_{y},p_{z}\right)$, satisfying:
\begin{equation}
\mathbf{p}=\frac{\hbar\omega}{c}\left(1,\hat{n}\right)=\frac{\hbar\omega}{c}\left(1,\cos\theta,\sin\theta,0\right),\label{eq:momentuminert}
\end{equation}
Now, let us obtain the photon's 4-momentum according to the accelerated
frame $\mathcal{A}$, that is $\bar{\mathbf{p}}=\left(p_{\tau},p_{\xi},\bar{p}_{y},\bar{p}_{z}\right).$
The dipersion relation (\ref{eq:disperssionrel}) is also satisfied
in $\mathcal{A}$, however, written in different coordinate, it becomes:
\begin{equation}
e^{2\alpha\xi}\bar{\omega}^{2}=e^{2\alpha\xi}k_{\xi}^{2}+k_{\bar{y}}^{2},\label{eq:energyconstraint-1-1}
\end{equation}
(notice that the dispersion relation comes from $p_{\mu}p^{\mu}=0,$
hence the metric components needs to be taken accounted). Since the
coordinate axis $\xi$ and $\bar{y}$ are also perpendicular to one
another, one could defined that $k_{\xi}=\bar{\omega}\cos\bar{\theta}$
and $k_{\bar{y}}=\bar{\omega}e^{\alpha\xi}\sin\bar{\theta}$. Notice
that $\bar{\theta}$ is the inclination angle according to $\mathcal{A}$.
With this, then the 4-momentum at the accelerated frame is:
\begin{equation}
\bar{\mathbf{p}}=\hbar\bar{\omega}\left(1,\hat{\bar{n}}\left[\xi,\tau\right]\right)=\hbar\bar{\omega}\left(1,\cos\bar{\theta},e^{\alpha\xi}\sin\bar{\theta},0\right).\label{eq:momentumrindler}
\end{equation}
$\bar{\omega}$ is the frequency of the photon, observed by $\mathcal{A}$.
Here, we assume that de Broglie postulate is still valid in $\mathcal{A},$
namely the relation $\bar{\mathbf{p}}\left[\tau,\xi\right]=\hbar\bar{\mathbf{k}}\left[\tau,\xi\right]$
is satisfied for every position in space and each instantaneous time.
One could refer to Appendix B  for a more detailed explanation on
the photon's momentum.

It needs to be kept in mind that there exists another solution to
(\ref{eq:disperssionrel}), namely: $\omega=-k,$ giving another possible
value for $k_{x}=-\omega\cos\theta$ and $k_{y}=-\omega\sin\theta$
in $\mathcal{O}$ and $k_{\xi}=-\bar{\omega}\cos\bar{\theta}$ and
$k_{\bar{y}}=-\bar{\omega}e^{\alpha\xi}\sin\bar{\theta}$ in $\mathcal{A}$.
However, since we are considering only the right Rindler wedge (RRW),
the reasonable case for an observer in RRW to receive the signal is
by the waveform that moves to the right (in $+x$ direction). For
an observer in the left Rindler wedge (LRW, which we will consider
in Section IV), we should consider the case where $\omega=-k.$ 

The (generalized) momentum is naturally a covector, however, in this
paper, we use its contravariant counterpart to be consistent with
the 4-velocity (\ref{eq:velocity5})-(\ref{eq:velocity6}). The entire
results do not depends on the choice of vector/covector for the derivation.
Since the 4-momentum is a 4-vector, the transformation between $\mathbf{p}$
and $\bar{\mathbf{p}}$ follows (\ref{eq:Rindler-1})-(\ref{eq:Rindler-2}),
where we write the cofficient of the transformation in terms of hyperbolic
functions:
\begin{equation}
\begin{array}{cc}
p_{\tau} & =e^{-\alpha\xi}\left(p_{t}\cosh\bar{\varsigma}-p_{x}\sinh\bar{\varsigma}\right),\\
p_{\xi} & =e^{-\alpha\xi}\left(p_{x}\cosh\bar{\varsigma}-p_{t}\sinh\bar{\varsigma}\right),
\end{array}\label{eq:Rindler-3}
\end{equation}
with their inverses:
\begin{equation}
\begin{array}{cc}
p_{t} & =e^{\alpha\xi}\left(p_{\tau}\cosh\bar{\varsigma}+p_{\xi}\sinh\bar{\varsigma}\right),\\
p_{x} & =e^{\alpha\xi}\left(p_{\xi}\cosh\bar{\varsigma}+p_{\tau}\sinh\bar{\varsigma}\right),
\end{array}\label{eq:invRindler-3}
\end{equation}
$\bar{p}_{y}=p_{y}$ and $\bar{p}_{z}=p_{z}.$ Moreover, by inserting
(\ref{eq:momentuminert})-(\ref{eq:momentumrindler}) to the transformation
(\ref{eq:Rindler-3}), we can obtain the 3 equivalent forms of the
relativistic Doppler effect in the accelerated frame as follows:
\begin{align}
\frac{\bar{\omega}}{\omega} & =e^{-\alpha\xi}\left(\cosh\bar{\varsigma}-\cos\theta\sinh\bar{\varsigma}\right),\label{eq:doppler1}\\
 & =e^{-\alpha\xi}\frac{\left(\cos\theta\cosh\bar{\varsigma}-\sinh\bar{\varsigma}\right)}{\cos\bar{\theta}},\label{eq:dope}\\
 & =\frac{\sin\theta}{\sin\bar{\theta}}\,e^{-\alpha\xi},\label{eq:sine}
\end{align}
together with their inverses:
\begin{align}
\frac{\omega}{\bar{\omega}} & =e^{\alpha\xi}\left(\cosh\bar{\varsigma}+\cos\bar{\theta}\sinh\bar{\varsigma}\right),\label{eq:doppler2}\\
 & =e^{\alpha\xi}\frac{\left(\cos\bar{\theta}\cosh\bar{\varsigma}+\sinh\bar{\varsigma}\right)}{\cos\theta}.\nonumber 
\end{align}
Notice that $\bar{\omega}=\bar{\omega}\left[\tau,\xi\right]$ is a
function of time and position in the Rindler coordinate. From (\ref{eq:doppler1})
and (\ref{eq:doppler2}), one could retrieve the aberration formula
(\ref{eq:aberrationrindler})-(\ref{eq:aberrationrindler2}).

Let us consider one of the light aberation formula (\ref{eq:vabe2}).
Differentiating (\ref{eq:vabe2}) with respect to any parameter $\lambda$
give:

\[
-\sin\theta\frac{d\theta}{d\lambda}=\frac{\sin\bar{\theta}\left(-\frac{d\bar{\theta}}{d\lambda}+\alpha\sin\bar{\theta}\frac{d\tau}{d\lambda}\right)}{\left(\cosh\bar{\varsigma}+\cos\bar{\theta}\sinh\bar{\varsigma}\right)^{2}},
\]
while using (\ref{eq:aberrationrindler2}), (\ref{eq:sine}), and
the chain rule $\frac{d\tau}{d\lambda}=\frac{d\tau}{d\bar{\theta}}\frac{d\bar{\theta}}{d\lambda}$,
gives:
\begin{equation}
\frac{d\theta}{d\bar{\theta}}=\frac{\bar{\omega}}{\omega}e^{\alpha\xi}\left(1-\frac{\alpha}{\dot{\bar{\theta}}}\sin\bar{\theta}\right),\label{eq:polarangle}
\end{equation}
with $\dot{\bar{\theta}}=\frac{d\bar{\theta}}{d\tau}$. $\dot{\bar{\theta}}$
is the rate of change of the inclination angle $\bar{\theta}$ as
observed by the accerelated observer $\mathcal{A}$. In $\mathcal{A}$,
the inclination angle $\bar{\theta}$ is time-dependent, satisfying
equation (\ref{eq:aberrationrindler}); differentiating (\ref{eq:aberrationrindler})
with respect to $\tau$, we obtain:
\begin{equation}
\dot{\bar{\theta}}=\frac{d\bar{\theta}}{d\tau}=\sin\bar{\theta},\label{eq:rateofchangeangle}
\end{equation}
and hence (\ref{eq:polarangle}) can be simplified into:
\begin{equation}
\frac{d\theta}{d\bar{\theta}}=\frac{\bar{\omega}}{\omega}e^{\alpha\xi}\left(1-\alpha\right).\label{eq:polarangle-1}
\end{equation}

Using  the transformation of the inclination (polar) angle (\ref{eq:polarangle}),
one could obtain the transformation for the solid angle $d\Omega=\sin\theta d\theta d\phi$
as follows:
\begin{equation}
\frac{d\Omega}{d\bar{\Omega}}=\left(\frac{\bar{\omega}}{\omega}\right)^{2}e^{2\alpha\xi}\left(1-\alpha\right),\label{eq:solidangleacc}
\end{equation}
where we use the Doppler effect (\ref{eq:sine}) and the fact that
the transformation of the azimuth angle satisfies $d\phi=d\bar{\phi}$. 

As explained in the preceeding subsection, observer $\mathcal{A}$
is accelerated along the $+x$ direction; causing it to receive only
photons approaching from that direction, namely, those with the $+x$
component in their velocity. Moreover, $\mathcal{A}$ is moving away
from the black-body source at rest with respect to $\mathcal{O}.$
As a consequence, the frequency of the photons received by the accelerated
frame $\mathcal{A}$ is redshifted by equation (\ref{eq:doppler1}).
However, in constrast with the inertial case presented in Appendix
B where the redshift remains constant, the redshift in the accelerated
frame $\mathcal{A}$ increases in time, with the wavelength of the
photons shift progressively towards the infrared range. To obtain
the blueshift case, one could invert the situation by flipping the
sign of $\tau$ (or $t$), effectively moving backward in time.

\subsection{Density of State in Accelerated Frame}

Another important parameter we must determined in the accelerated
frame $\mathcal{A}$ is the modes distribution of the photons in the
cavity. To determine how this quantity transform from the inertial
to accelerated frame, first we need to determine transformation of
the phase-space volume. In the inertial frame $\mathcal{O}$ and the
accelerated frame $\mathcal{A}$, the infinitesimal 3-volume elements
are defined as, respectively:
\begin{align}
d^{3}\mathbf{x} & =dx\wedge dy\wedge dz,\label{eq:volumeelement}\\
d^{3}\mathbf{\bar{x}} & =d\xi\wedge d\bar{y}\wedge d\bar{z}.\nonumber 
\end{align}
To obtain the density of states of our black-body system, we need
to calculate how much modes are inside a volume element. Let the finite
volume element in $\mathcal{O}$ be $\Delta x\Delta y\Delta z$. The
spatial length $\Delta x$ is defined by 2 simultaneous events $\mathbf{p}=\left(t_{i},x_{i}\right)$
and $\mathbf{q}=\left(t_{f},x_{f}\right)$ along the $x$-axis where
$t_{i}=t_{f}=0$, $x_{i}=0$ and $x_{f}=\ell$. According to $\mathcal{O},$
the length between these 2 event is $\Delta x=x_{f}-x_{i}=\ell.$
For an observer at $\mathcal{A},$ whose accelerated in the $x$-direction
with respect to $\mathcal{O},$ the 2 events $\mathbf{p}=\left(\tau_{i},\xi_{i}\right)$
and $\mathbf{q}=\left(\tau_{f},\xi_{f}\right)$ are not simultaneous,
but are separated by a time interval $\Delta\tau=\tau_{f}-\tau_{i}$.
Let us labeled the spatial length of these 2 events in $\mathcal{A}$
as $\Delta\xi=\xi_{f}-\xi{}_{i},$where $\tau_{f},\tau_{i},\xi{}_{i},\xi_{f},$
could be obtained from (\ref{eq:Rindler}). Taking the infinitesimal
limit $\Delta\xi\rightarrow d\xi$, the transformation of the infinitesimal
spatial length satisfies (\ref{eq:Rindler-1}). Notice that the measurement
of length in $\mathcal{O}$ is obtained from 2 simultaneous event,
hence in $\mathcal{O}$, $dt=0;$ this is not the case in $\mathcal{A}.$
Therefore, the (infinitesimal) spatial length of $\mathbf{p},\mathbf{q}$
in $\mathcal{A}$ is:
\begin{equation}
d\xi=e^{-\alpha\xi}\cosh\bar{\varsigma}\,dx=\frac{x}{\alpha\left(x^{2}-t^{2}\right)}\,dx.\label{eq:if}
\end{equation}
Furthermore, the infinitesimal volume element $dx\wedge dy\wedge dz$
is perceived by an observer in $\mathcal{A}$ as the volume swept
by the plane $d\bar{y}\wedge d\bar{z}=dy\wedge dz$ from $\xi{}_{i}$
to $\xi_{f},$ along an infinitesimal time interval $d\tau.$ This
is the physical interpretation of $d^{3}\mathbf{\bar{x}}=d\xi\wedge d\bar{y}\wedge d\bar{z},$
see also discussion on Appendix B. Inserting (\ref{eq:if}) to (\ref{eq:volumeelement})
gives the volume transformation:
\begin{equation}
d^{3}\mathbf{\bar{x}}=e^{-\alpha\xi}\cosh\bar{\varsigma}\,d^{3}\mathbf{x}=\frac{x}{\alpha\left(x^{2}-t^{2}\right)}\,d^{3}\mathbf{x}.\label{eq:volumecontractionrindler}
\end{equation}

For the next step, we need to determine the infinitesimal volume element
in the momentum space, this could be obtained from (\ref{eq:momentuminert})-(\ref{eq:momentumrindler});
in the inertial frame $\mathcal{O}$ and the accelerated frame $\mathcal{A}$,
they are, respectively:
\begin{align}
d^{3}\mathbf{p} & =dp_{x}\wedge dp_{y}\wedge dp_{z},\label{eq:volumeelement-1}\\
d^{3}\mathbf{\bar{p}} & =dp_{\xi}\wedge d\bar{p}_{y}\wedge d\bar{p}_{z}.\nonumber 
\end{align}
The 4-momentum is subjected to the energy-momentum constraint, see
Appendix B:
\begin{equation}
\left|^{4}\mathbf{p}\right|=p_{\mu}p^{\mu}=-p_{t}^{2}+p_{x}^{2}+p_{y}^{2}+p_{z}^{2}=m^{2}c^{2}.\label{eq:momconstraint}
\end{equation}
Taking the differential of (\ref{eq:momconstraint}) gives:
\begin{equation}
dp_{t}=\frac{1}{p_{t}}\left(p_{x}dp_{x}+p_{y}dp_{y}+p_{z}dp_{z}\right).\label{eq:momentumconstraint}
\end{equation}
Furthermore, using (\ref{eq:invRindler}), one could show that:
\begin{equation}
x^{2}-t^{2}=\frac{1}{\alpha^{2}}e^{2\alpha\xi},\label{eq:penting}
\end{equation}
so that the transformation of the 4-momentum (\ref{eq:Rindler-3})
can be written as follows:
\begin{align}
p_{\tau} & =\frac{1}{\alpha\left(x^{2}-t^{2}\right)}\left(xp_{t}-tp_{x}\right),\label{eq:momentum2}\\
p_{\xi} & =\frac{1}{\alpha\left(x^{2}-t^{2}\right)}\left(xp_{x}-tp_{t}\right),\nonumber 
\end{align}
with $\bar{p}_{y}=p_{y}$ and $\bar{p}_{z}=p_{z}.$ Differentiating
(\ref{eq:momentum2}) with any parameter will give the infinitesimal
version of 4-momentum transformation as follows: \begin{widetext}
\begin{align}
dp_{\tau} & =\frac{1}{\alpha\left(x^{2}-t^{2}\right)}\left[\left(2t\left[\frac{xp_{t}-tp_{x}}{x^{2}-t^{2}}\right]-p_{x}\right)dt+\left(p_{t}-2x\left[\frac{xp_{t}-tp_{x}}{x^{2}-t^{2}}\right]\right)dx+xdp_{t}-tdp_{x}\right],\label{eq:momentum2-1}\\
dp_{\xi} & =\frac{1}{\alpha\left(x^{2}-t^{2}\right)}\left[\left(2t\left[\frac{xp_{x}-tp_{t}}{x^{2}-t^{2}}\right]-p_{t}\right)dt+\left(p_{x}-2x\left[\frac{xp_{x}-tp_{t}}{x^{2}-t^{2}}\right]\right)dx+xdp_{x}-tdp_{t}\right].\nonumber 
\end{align}
 \end{widetext} Note  that in equation (\ref{eq:momentum2-1}), there
exist infinitesimal coordinates components, namely $dt$ and $dx$.
This is due to the fact that the 4-momentum transformation from inertial
to accelerated frame (\ref{eq:momentum2}) varies with coordinates
$\left(x,t\right)$, in constrast to the transformation between inertial
frames (\ref{eq:Lorentz-4}) in Appendix B, which is independent from
the temporal and spatial coordinates.

Similar to the inertial case in Appendix B, using the hypersurface
constraint $dt=0$ and inserting the infinitesimal momentum constraint
(\ref{eq:momentumconstraint}), (\ref{eq:momentum2-1}) becomes: \begin{widetext}
\begin{align}
dp_{\tau} & =\frac{1}{\alpha\left(x^{2}-t^{2}\right)}\left[\left(p_{t}-2x\left[\frac{xp_{t}-tp_{x}}{x^{2}-t^{2}}\right]\right)dx+\left(x\frac{p_{x}}{p_{t}}-t\right)dp_{x}+x\frac{1}{p_{t}}\left(p_{y}dp_{y}+p_{z}dp_{z}\right)\right],\label{eq:momentum3}\\
dp_{\xi} & =\frac{1}{\alpha\left(x^{2}-t^{2}\right)}\left[\left(p_{x}-2x\left[\frac{xp_{x}-tp_{t}}{x^{2}-t^{2}}\right]\right)dx+\left(x-t\frac{p_{x}}{p_{t}}\right)dp_{x}-t\frac{1}{p_{t}}\left(p_{y}dp_{y}+p_{z}dp_{z}\right)\right],\nonumber 
\end{align}
 \end{widetext} $d\bar{p}_{y}=dp_{y},$ and $d\bar{p}_{z}=dp_{z}$.
Now, one can construct altogether the infinitesimal phase-space volume
element in frame $\mathcal{A}$ as follows:
\[
d^{3}\bar{\mathbf{x}}\wedge d^{3}\bar{\mathbf{p}}=d\xi\wedge d\bar{y}\wedge d\bar{z}\wedge dp_{\xi}\wedge d\bar{p}_{y}\wedge d\bar{p}_{z}.
\]
 Inserting (\ref{eq:volumecontractionrindler}) and (\ref{eq:momentum3})
into the equation above, we obtain the transformation of the phase-space
volume element from the inertial frame $\mathcal{O}$ to the accelerated
frame $\mathcal{A}$:
\begin{equation}
d^{3}\bar{\mathbf{x}}\wedge d^{3}\bar{\mathbf{p}}=\frac{x\left(x-t\frac{p_{x}}{p_{t}}\right)}{\alpha^{2}\left(x^{2}-t^{2}\right)^{2}}\,d^{3}\mathbf{x}\wedge d^{3}\mathbf{p}.\label{eq:phasespacerindler}
\end{equation}
Note that by the definition of the wedge product $\wedge,$ the terms
containing equivalent components of differential forms, i.e. $dx\wedge dx$,
vanishes. Using (\ref{eq:penting}) and then (\ref{eq:momentuminert}),
one could simplify (\ref{eq:phasespacerindler}) into:
\begin{equation}
\frac{d^{3}\bar{\mathbf{x}}d^{3}\bar{\mathbf{p}}}{d^{3}\mathbf{x}d^{3}\mathbf{p}}=\left(1-\frac{p_{x}}{p_{t}}\tanh\bar{\varsigma}\right)e^{-2\alpha\xi}\cosh^{2}\bar{\varsigma}.\label{eq:transfrmom}
\end{equation}
 One the other hand, we have, from (\ref{eq:Rindler-3}):
\[
\frac{p_{\tau}}{p_{t}}=\left(1-\frac{p_{x}}{p_{t}}\tanh\bar{\varsigma}\right)e^{-\alpha\xi}\cosh\bar{\varsigma},
\]
and therefore, using (\ref{eq:momentuminert})-(\ref{eq:momentumrindler}):
\begin{equation}
\frac{d^{3}\bar{\mathbf{x}}d^{3}\bar{\mathbf{p}}}{d^{3}\mathbf{x}d^{3}\mathbf{p}}=\frac{p_{\tau}}{p_{t}}e^{-\alpha\xi}\cosh\bar{\varsigma}=\frac{\bar{\omega}}{\omega}e^{-\alpha\xi}\cosh\bar{\varsigma},\label{eq:momemntumtransform}
\end{equation}
where we write the transformation (\ref{eq:transfrmom}) in terms
of the Doppler factor $\frac{\bar{\omega}}{\omega}$. (\ref{eq:momemntumtransform})
is the relation between the phase-space volume element in $\mathcal{O}$
and $\mathcal{A}.$

Let us define the relativistic distribution function, or \textit{density
of state} $f\left(\mathbf{x},\mathbf{p}\right)$ as the number of
world-lines that cross the phase-space, i.e, the states $dN$, per
phase-space volume element \citep{Liboff}:
\begin{equation}
f\left(\mathbf{x},\mathbf{p}\right)=\frac{dN}{d^{3}\mathbf{x}d^{3}\mathbf{p}}.\label{eq:distrirel}
\end{equation}
We assume that the density of state $f\left(\mathbf{x},\mathbf{p}\right)$
is invariant under coordinate transformation, namely:
\[
\frac{dN}{d^{3}\mathbf{x}d^{3}\mathbf{p}}=\frac{d\bar{N}}{d^{3}\mathbf{\bar{x}}d^{3}\mathbf{\bar{p}}},
\]
with $d\bar{N}$ is the number of states per phase-space volume $d^{3}\mathbf{\bar{x}}\,d^{3}\mathbf{\bar{p}}$
of the accelerated frame $\mathcal{A}$. This is a reasonable assumption,
and for a detailed explanation, one could refer to Appendix B. Some
previous works in the existing literature have assume the invariance
of the number of states $dN$ instead of $f\left(\mathbf{x},\mathbf{p}\right)$,
however, this assumption is not suitable for our case because our
treatment in this works relies only on a coordinate transformations
at some part of the Minkowski space, i.e., the Rindler wedge. While
these transformations leave the world-lines invariant, they alter
the unit volume due to the coordinate transformation, leading to a
different count of world-lines crossing the unit volume. Similar arguments
are implicitly employed in \citep{Peebles,Heer,Henry} as well.

By the invariance of the density of state under (infinitesimal) coordinate
transformation, one could obtain the relation between the number of
states/modes in inertial frame $\mathcal{O}$ and the accelerated
frame $\mathcal{A}$ as follows:
\begin{equation}
\frac{d\bar{N}}{dN}=\frac{d^{3}\bar{\mathbf{x}}d^{3}\bar{\mathbf{p}}}{d^{3}\mathbf{x}d^{3}\mathbf{p}}=\frac{\bar{\omega}}{\omega}e^{-\alpha\xi}\cosh\bar{\varsigma}.\label{eq:numberacc}
\end{equation}

\subsection{Black-body Radiation in Accelerated Frame}

We already had all the Rindler-transformed variables to derive the
Planck's law in a uniformly-accelerated frame. Similar to the inertial
frame case that we derived in Appendix B, in the rest frame $\mathcal{O}$,
we use the Planck's law version of equation (\ref{eq:Plancklawnew-2})
as follows:
\begin{equation}
dN=\frac{\omega^{2}}{4\pi^{3}c^{3}}\frac{1}{e^{\nicefrac{w\omega}{\omega_{\mathrm{max}}}}-1}d\omega d\Omega dV,\label{eq:Plancklawnew}
\end{equation}
where the term containing the temperature of the black-body, i.e.,
$k_{B}T$, is replaced by $\omega_{\mathrm{max}}$, the angular frequency
that maximize the radiation energy $E$ of the black-body. This is
possible by the Wien's displacement law, see a detailed explanation
on the Appendix A. The reason for this substitution is to avoid the
problem of the temperature in moving bodies. As explained in the Introduction,
currently, there is no consensus regarding the transformation of temperature
in moving bodies \citep{Farias}. However, we convincingly agree on
the transformation of the black-body's frequency, hence, if $\omega$
satisfies the relativistic Doppler effect, then so does the $\omega_{\mathrm{max}}$.
The dimensionless constant $w$ in (\ref{eq:Plancklawnew}) is the
Wien coefficient, obtained from solving the non-linear equation that
arise from the maximization of energy with respect to $\omega$. One
could refer to Appendix B for a detailed derivation.

In this section, our objective is to ascertain the validity of the
distribution (\ref{eq:Plancklawnew}) in the accelerated frame $\mathcal{A}$.
Inserting the relativistic Doppler effect (\ref{eq:doppler2}), the
transformation of solid angle (\ref{eq:solidangleacc}), the volume
contraction (\ref{eq:volumecontractionrindler}), and the transformation
of the number of modes distribution (\ref{eq:numberacc}) to (\ref{eq:Plancklawnew}),
we obtain:
\begin{equation}
d\bar{N}=\left(1-\alpha\right)\frac{e^{2\alpha\xi}}{4\pi^{3}c^{3}}\frac{\bar{\omega}^{2}}{e^{\nicefrac{w\bar{\omega}}{\overline{\omega_{\mathrm{max}}}}}-1}d\bar{\omega}d\bar{\Omega}d\bar{V}.\label{eq:planckaccel}
\end{equation}
One can observe that in the right hand side of (\ref{eq:planckaccel})
there is a scale factor:
\begin{equation}
\varepsilon\left[\alpha,\xi\right]=e^{2\alpha\xi}\left(1-\alpha\right),\label{eq:emmissivity}
\end{equation}
that prevents the distribution (\ref{eq:planckaccel}) to be the original
Planckian (\ref{eq:Plancklawnew}). The scale factor depends on the
acceleration $\alpha$ and the spatial coordinate $\xi$. For the
case with $\alpha=0$ where the acceleration vanishes, the distribution
becomes Planckian as in (\ref{eq:Plancklawnew}). The fundamental
distinction between the distribution (\ref{eq:planckaccel}) and the
distribution (\ref{eq:Plancklawnewmoving-1}) in the moving inertial
frame lies in the fact that all the variables in the (\ref{eq:planckaccel})
are coordinate-dependent, namely, they vary with $\tau$ and $\xi$,
which represent the proper time and position in $\mathcal{A}$. This
is due to the fact that the transformations of these variables, namely
(\ref{eq:doppler2}), (\ref{eq:solidangleacc}), (\ref{eq:volumecontractionrindler}),
and (\ref{eq:numberacc}) are coordinate-dependent. However, since
we are considering all the possible value of frequencies from $0\leq\omega<\infty$,
the time dependence of $\bar{\omega}$ does not explicitly shown in
the law. Consequently, the 'black-body' spectrum (\ref{eq:planckaccel})
in the accelerated frame $\mathcal{A}$ experiences a redshift that
increase over time and varying with position, where the wavelength
of the photon increases toward the infrared direction. This behavior
distinguishes (\ref{eq:planckaccel}) from the black-body spectrum
in the inertial frames, where the spectrum shifts similarly occur,
but the magnitudes of these shifts remain constant, as indicated by
equation (\ref{eq:doppler}). 

Similar with the inertial case in Appendix B, $\overline{\omega_{\mathrm{max}}}$
is the ``maximal'' frequency of the distribution (\ref{eq:Plancklawnew})
of a black-body in the rest frame, viewed by the accelerated observer
$\mathcal{A}$. Let us check if $\overline{\omega_{\mathrm{max}}}$
also maximizes the energy in distribution (\ref{eq:planckaccel}).
By multiplying the left and right hand side of (\ref{eq:planckaccel})
with $\hbar\bar{\omega},$ one obtains:
\begin{equation}
d\bar{E}=\underset{\bar{\rho}{}_{E}\left[\bar{\omega}\right]}{\underbrace{\frac{\varepsilon}{4\pi^{3}c^{3}}\frac{\hbar\bar{\omega}^{3}}{e^{\nicefrac{w\bar{\omega}}{\overline{\omega_{\mathrm{max}}}}}-1}}}d\bar{\omega}d\bar{\Omega}d\bar{V}.\label{eq:Plancklawnewmoving-1}
\end{equation}
$\bar{E}$ is the energy of modes with frequency $\bar{\omega}$ observed
in the accelerated frame $\mathcal{A}$. The value of $\bar{\omega}$
that maximize $\bar{E}$ can be obtained by solving the following
equation: \begin{widetext}
\begin{equation}
\frac{d\bar{\rho}{}_{E}}{d\bar{\omega}}=\frac{\varepsilon}{4\pi^{3}c^{3}}\frac{\hbar\bar{\omega}^{2}}{\exp\left(w\frac{\bar{\omega}}{\overline{\omega_{\mathrm{max}}}}\right)-1}\left(3-w\frac{\bar{\omega}}{\overline{\omega_{\mathrm{max}}}}\frac{\exp\left(w\frac{\bar{\omega}}{\overline{\omega_{\mathrm{max}}}}\right)}{\exp\left(w\frac{\bar{\omega}}{\overline{\omega_{\mathrm{max}}}}\right)-1}\right)=0.\label{eq:dplanckacc}
\end{equation}
 \end{widetext} Note that the scale factor $\varepsilon$ does not
affect the equation we need to solve. Let us define $\bar{w}=w\frac{\bar{\omega}}{\overline{\omega_{\mathrm{max}}}}$,
the equation from (\ref{eq:dplanckacc}) that we need to solve becomes:
\begin{equation}
\bar{w}=3\left(1-e^{-\bar{w}}\right),\label{eq:Lambertmov}
\end{equation}
which is the same as in the inertial cases, namely (\ref{eq:Lambert})
and (\ref{eq:Lambertmov-1}). The solution is then similar:
\[
\bar{w}=3+W_{0}\left(-3e^{-3}\right),
\]
with $W_{0}$ is the Lambert $W$-function \citep{Euler}. Hence,
the Wien's coefficient is invariant, even under the transformation
from the inertial to accelerated frame: $\bar{w}=w$. Suppose $\bar{\omega}=\bar{\omega}_{\mathrm{max}}$
is the solution to (\ref{eq:dplanckacc}), therefore: 
\[
\frac{\bar{w}}{w}=\frac{\bar{\omega}_{\mathrm{max}}}{\overline{\omega_{\mathrm{max}}}}=1,
\]
or simply $\bar{\omega}_{\mathrm{max}}=\overline{\omega_{\mathrm{max}}}$:
the transformation (\ref{eq:doppler1}) sends ``maximal'' frequency
in an inertial frame to a ``maximal'' frequency in an accelerated
frame at an instantaneous time. (\ref{eq:planckaccel}) becomes:
\begin{equation}
d\bar{N}=\frac{\varepsilon}{4\pi^{3}c^{3}}\frac{\bar{\omega}^{2}}{e^{\nicefrac{w\bar{\omega}}{\bar{\omega}_{\mathrm{max}}}}-1}d\bar{\omega}d\bar{\Omega}d\bar{V}.\label{eq:planckaccel-1}
\end{equation}

As previously mentioned, the presence of the scale factor $\varepsilon$
introduces deviations from the Planckian distribution of the black-body
radiation in the accelerated frame $\mathcal{A}$. For an observer
$\mathcal{A}$ moving along $\xi=0$, it will observe a Planckian
distribution but with a redshift increasing over time. For another
accelerated observer moving along a constant $\xi>0$ , it will experience
a similar increasing redshift, but now the distribution will be scaled
by a factor of $e^{2\alpha\xi}\left(1-\alpha\right)$. Such an observer
will interpret this as a change in the emissivity of the black-body
source; this facts will be analyzed in detail in the next section.

It needs to be kept in mind that the Planck formula (\ref{eq:planckaccel-1})
applies in the right Rindler wedge (RRW). With a similar derivation,
one could show that the same formula is also valid for the left Rindler
wedge, but with the Doppler formula (\ref{eq:doppler1})-(\ref{eq:dope})
changed into:
\begin{align}
\frac{\bar{\omega}}{\omega} & =e^{-\alpha\xi}\left(\cosh\bar{\varsigma}+\cos\theta\sinh\bar{\varsigma}\right),\label{eq:doppler1LRW}\\
 & =e^{-\alpha\xi}\frac{\left(\cos\theta\cosh\bar{\varsigma}+\sinh\bar{\varsigma}\right)}{\cos\bar{\theta}}.\label{eq:dope-1}
\end{align}
This is due to the fact that there exists another solution to the
dispersion relation (\ref{eq:disperssionrel}), namely $\omega=-k.$ 

\section{Physics in Rindler Coordinate}

\subsection{Acceleration and Deceleration}

The \textquotedbl Planck's law\textquotedbl{} (\ref{eq:planckaccel-1})
describes the frequency spectrum measured by an observer $\mathcal{A}$
whose accelerating away from the radiation source along the $+x$
direction, as described in Section II. This observer moves with an
initial velocity $\beta=\tanh\varsigma$ and undergoes an acceleration
with magnitude $a$. Let us add three possible scenarios besides the
ones we had considered in the previous section. First, note that the
origin of the rest frame $\mathcal{O}$ does not coincide to the origin
of frame $\mathcal{\mathcal{A}},$ namely, the point $\left(t,x\right)=\left(0,0\right)$
at $\mathcal{O}$ is equivalent to point of $\left(\tau,\xi\right)=\left(-\frac{\varsigma}{\alpha},-\infty\right)$
at $\mathcal{\mathcal{A}},$ while the point of $\left(\tau,\xi\right)=\left(0,0\right)$
at $\mathcal{\mathcal{A}},$ is the point of $\left(t,x\right)=\frac{1}{\alpha}\left(\sinh\varsigma,\cosh\varsigma\right)$
at $\mathcal{O},$ see FIG. 2 and FIG. 3. 
\begin{figure}
\begin{centering}
\includegraphics[scale=0.65]{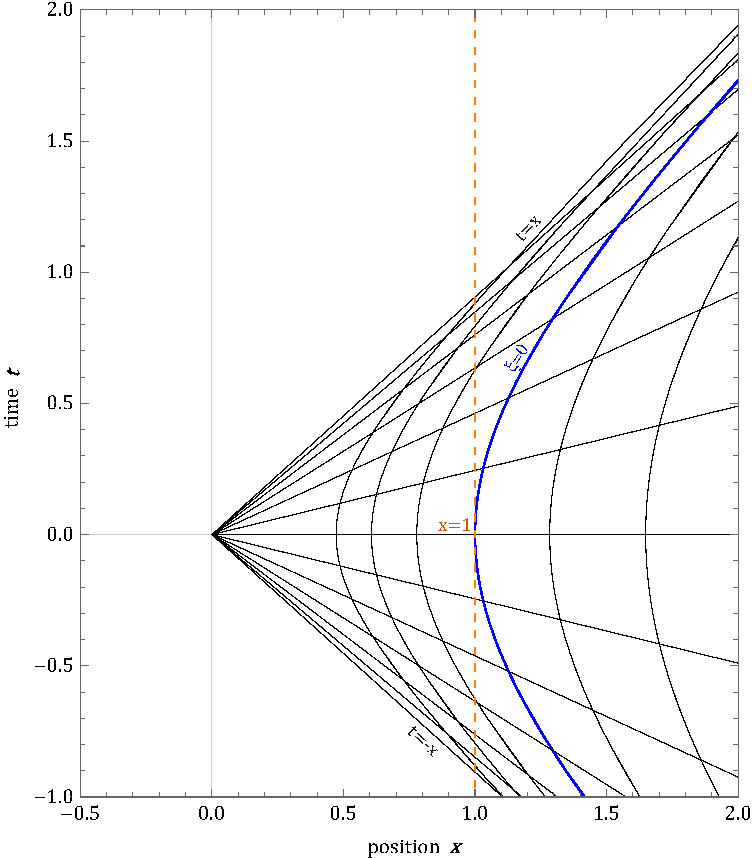}
\par\end{centering}
\caption{The accelerated away case as observed by the rest frame $\mathcal{O}$.
Let us set $\alpha=1$ and $\varsigma=0$, so that the origin of $\mathcal{A}$
is at $\left(t,x\right)=\left(0,1\right)$. The source is located
at $x=1$ (or $\xi=0$), at rest with respect to $\mathcal{O}$, and
its worldline with respect to $\mathcal{O}$ is described by the orange-dashed
line. $\mathcal{A}$ is accelerated with respect to $\mathcal{O}$,
hence its wordline is described by the blue hyperboloid. Note that
the distance between the source and $\mathcal{A}$ increases, and
so does the time interval to receive the signal. The source signal
at the horizon (and above) will never reach $\mathcal{A}$ since it
needs infinite time to reach the observer.}
\end{figure}
 
\begin{figure}
\centering{}\includegraphics[scale=0.65]{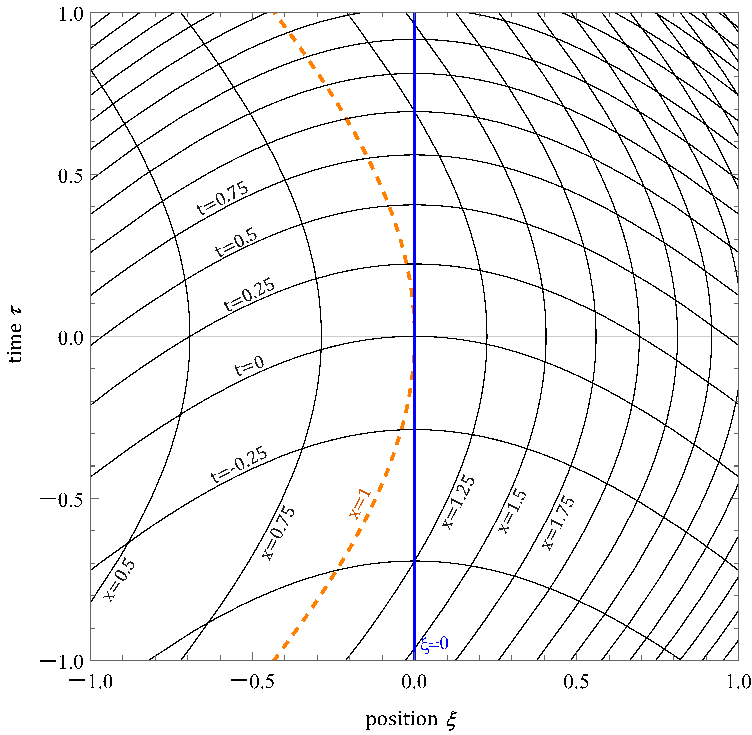}\caption{The accelerated away case as observed by the rest frame $\mathcal{A}$.
The origin of $\mathcal{O}$ is at $\left(\tau,\xi\right)=\left(0,-\infty\right)$
(not shown). Now the worldline of $\mathcal{A}$ is constant in the
Rindler coordinate, described by the blue line $\xi=0$. The source,
located at $\xi=0$ when $\tau=0$, is moving away from $\mathcal{A},$
described by the orange-dashed hyperboloid. This is exactly the same
situation as in FIG. 2 but described in the Rindler coordinate.}
\end{figure}

To simplify the analysis, let us set the black-body source to be located
at any point in the range of $0\leq x\leq\frac{1}{\alpha}\cosh\varsigma$
or $\frac{1}{\alpha}\cosh\varsigma<x\leq\infty$, depending on the
scenarios. With respect to $\mathcal{O}$, let frame $\mathcal{A}$
have an initial point at $\left(t,x\right)=\frac{1}{\alpha}\left(\sinh\varsigma,\cosh\varsigma\right)$,
that means the clock in $\mathcal{A}$ is slightly delayed by the
amount of $\frac{1}{\alpha}\sinh\varsigma$. Then $\mathcal{A}$ moves
with the time-dependent rapidity $\bar{\varsigma}\left[\tau\right]$,
which could satisfy one of the four following case:
\begin{equation}
\bar{\varsigma}\left[\tau\right]=\pm\varsigma\pm\alpha\tau,\label{eq:awaytoward}
\end{equation}
noting that the initial rapidity $\varsigma\geq0$. The sign $\pm$
in front of the initial rapidity $\varsigma$ depends on the rate
of change of the position of observer $\mathcal{A}$ with respect
to the source; it is positive if their distance increases in time,
and negative if it decreases. Meanwhile, the sign $\pm$ in front
of $\alpha\tau$ depends on the rate of change of $\varsigma$. If
it increases in time (accelerating), then it has the same sign with
$\varsigma$, and vice versa. Let us list all the possible cases.
The first case, where $\bar{\varsigma}\left[\tau\right]=\varsigma+\alpha\tau$,
corresponds to the case described in Section II, used to derive the
Planck's law (\ref{eq:planckaccel-1}), see FIG. 4. 
\begin{figure}
\begin{centering}
\includegraphics[scale=0.65]{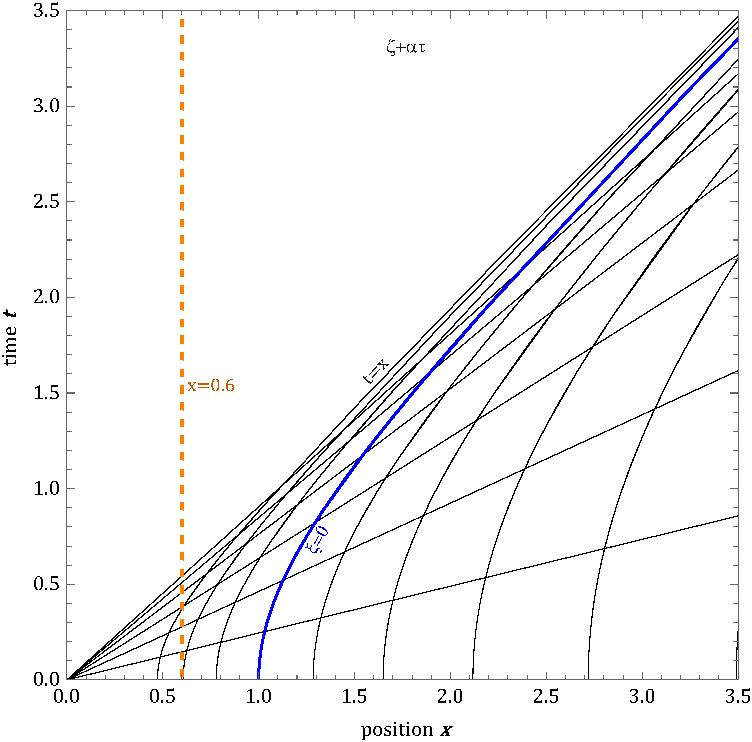}
\par\end{centering}
\caption{Case 1: The ``accelerating away'' case could be realized in the
right Rindler wedge by locating the source (orange dashed-line) at
$0\protect\leq x\protect\leq\frac{1}{\alpha}\cosh\varsigma$, where
$\frac{1}{\alpha}\cosh\varsigma$ is the origin of $\mathcal{A}.$
In this case, the source is located at $x=0.6$ and the origin of
$\mathcal{A}$ is at $x=1,$ given $\alpha=1$ and $\varsigma=0$.
The trajectory of observer $\mathcal{A}$ is described by the blue
curve.}
\end{figure}
 The second case, where $\bar{\varsigma}\left[\tau\right]=\varsigma-\alpha\tau$,
describes an observer decelerating \textit{away} from the source;
it moves in the $\ensuremath{+x}$ direction, but the speed is decreasing
in time; we will only consider this case until the observer reach
zero velocity as a result of deceleration. See FIG. 5. 
\begin{figure}
\begin{centering}
\includegraphics[scale=0.65]{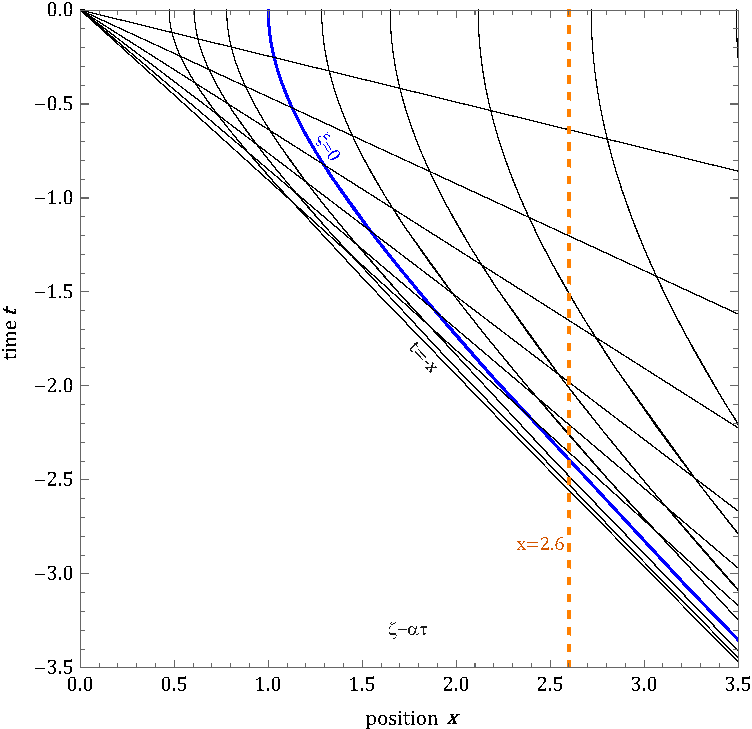}
\par\end{centering}
\caption{Case 2: The second case could be similarly realized in the right Rindler
wedge by locating the source at $\frac{1}{\alpha}\cosh\varsigma<x\protect\leq\infty$,
in this case $x=2.6$. The observer's trajectory coincides with the
source at some time in the past $\tau_{i}<0$. At $\tau_{i},$ the
(initial) velocity of $\mathcal{A}$ is $\varsigma$; which decreases
as time progresses by the amount of $-\alpha\tau$, and eventually,
reaching zero at $\tau_{f}=0$. At this time, the observer is at a
finite distance from the source. The collection of events in the time
interval of $\tau_{i}\protect\leq\tau\protect\leq\tau_{f}$ is considered
as the ``decelerating away'' case.}
\end{figure}
 The third case with $\bar{\varsigma}\left[\tau\right]=-\varsigma+\alpha\tau$
describes an observer that is decelerating \textit{towards} the source,
with a decreasing speed. Similar to the second case, we will only
consider the case until the velocity is zero, see FIG. 6. 
\begin{figure}
\begin{centering}
\includegraphics[scale=0.65]{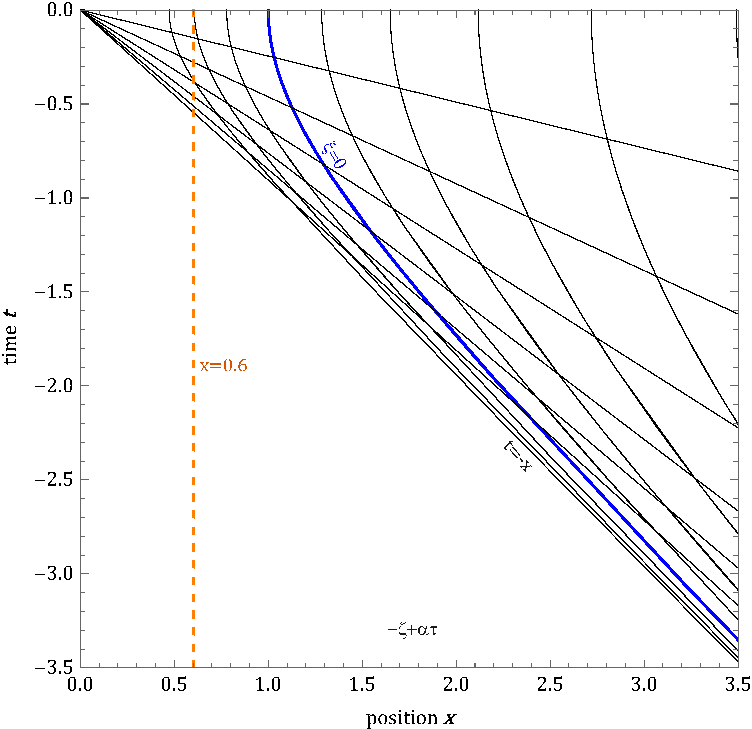}
\par\end{centering}
\caption{Case 3: To describe the ``decelerated towards'' case in the right
Rindler wedge, the source is located at $0\protect\leq x\protect\leq\frac{1}{\alpha}\cosh\varsigma$,
in this case $x=0.6$. The observer approaches the source from $x=\infty$
with an initial velocity $-\varsigma$, starts to decelerate, and
then stops at $\tau=0$.}
\end{figure}
 The last case with $\bar{\varsigma}\left[\tau\right]=-\varsigma-\alpha\tau$
describes an observer accelerating \textit{towards} the source, see
FIG. 7. 
\begin{figure}
\begin{centering}
\includegraphics[scale=0.65]{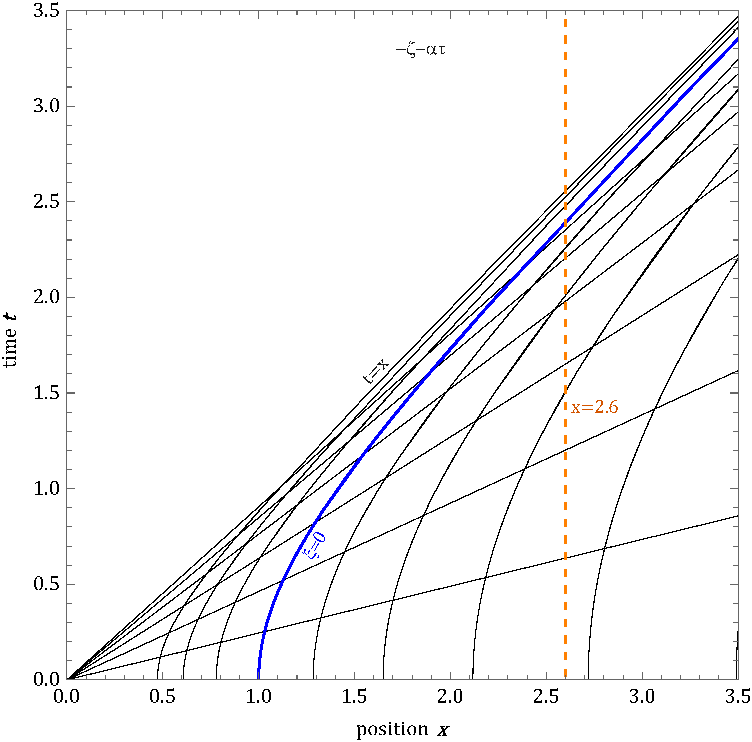}
\par\end{centering}
\caption{Case 4: To describe the ``accelerated towards'' in the right Rindler
wedge, the source is located at $\frac{1}{\alpha}\cosh\varsigma<x\protect\leq\infty$,
namely $x=2.6$. The observer approaches the source from the origin
of $\mathcal{A}$ with an increasing velocity and then coincides with
the source at a time $\tau>0$ in the future.}
\end{figure}

The concept of acceleration and deceleration is formally defined as
the rate of change of the speed, specifically, an acceleration if
$\frac{d\left|\bar{\varsigma}\right|}{d\tau}>0$ and deceleration
if $\frac{d\left|\bar{\varsigma}\right|}{d\tau}<0$. It is also important
to highlight that the notion of acceleration/deceleration do not necessarily
correlate to redshift/blueshift, as elaborated in Subsection III C.

\subsection{The Relativistic Beaming}

The alteration of the inclination angle in the moving frame leads
to a peculiar phenomenon known as relativistic beaming where, in moving
a frame, the rays of light shows a tendency to either converge or
diverge in the direction of motion, depending on whether the observer
is approaching or receding away from the source \citep{Cohen}. For
a frame $\mathcal{O}'$ moving away with constant velocity of $\beta$
, the inclination angle $\theta$ will transform by (\ref{eq:aberration-2}).
For the case where the observer is moving towards the source, one
could flip the sign in front $\beta$ to become positive. The photons
that reach the observer will have the inclination angle within the
range of $-\frac{\pi}{2}\leq\theta\leq\frac{\pi}{2}$. It can be demonstrated
that for the ``moving towards'' case ($+\beta$), the condition
$1\leq\gamma\left(1+\beta\cos\theta\right)\leq\infty$ holds, implying
$\left|\theta'\right|\leq\left|\theta\right|$. It implies that for
an observer approaching the source, the light rays tend to converge
to the direction of motion, see FIG. 8. 
\begin{figure}
\centering{}\includegraphics[scale=0.65]{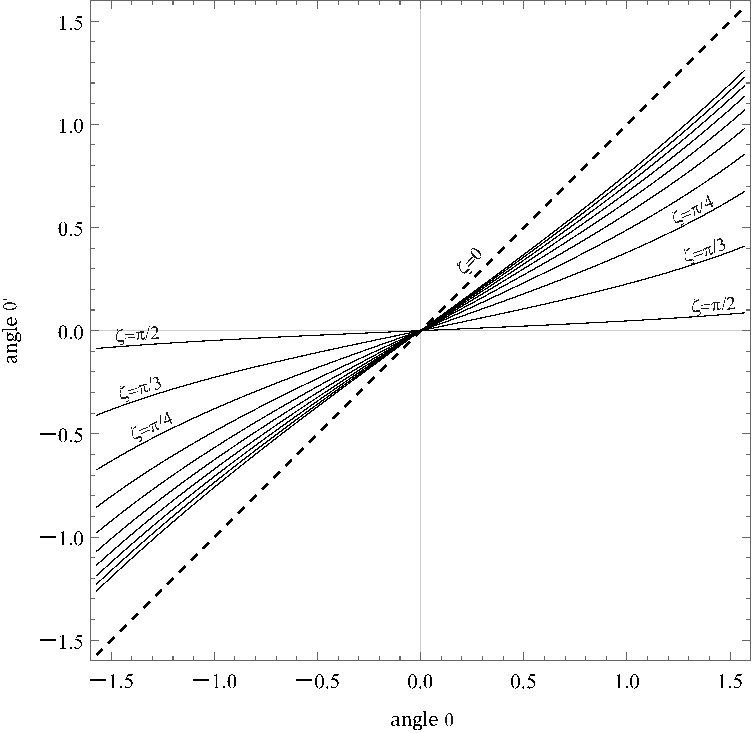}\caption{The alteration of the inclination angle on a frame approaching the
source with velocity $\varsigma$. The slopes of the collection of
curves $\theta'\left[\theta\right]$ never exceed 1, except for the
case where $\varsigma=0$, described by the dashed-line. As the rapidity
$\varsigma$ increases, the slope of the curve decreases, hence $\left|\theta'\right|\protect\leq\left|\theta\right|.$ }
\end{figure}
 For the ``moving away'' case ($-\beta$), there are 2 conditions
that affect the light rays bending. First, if $1\leq\gamma\left(1-\beta\cos\theta\right)\leq\infty$
holds, then $\left|\theta'\right|\leq\left|\theta\right|$, implying
that in this interval, the light rays converge. Second, if the condition
$0\leq\gamma\left(1-\beta\cos\theta\right)\leq1$ holds, the light
rays diverge in the direction of motion, namely $\left|\theta'\right|\geq\left|\theta\right|$.
This case occurs when $\cos\theta\leq\frac{\beta}{\beta-2}$, see
FIG. 9. 
\begin{figure}
\centering{}\includegraphics[scale=0.65]{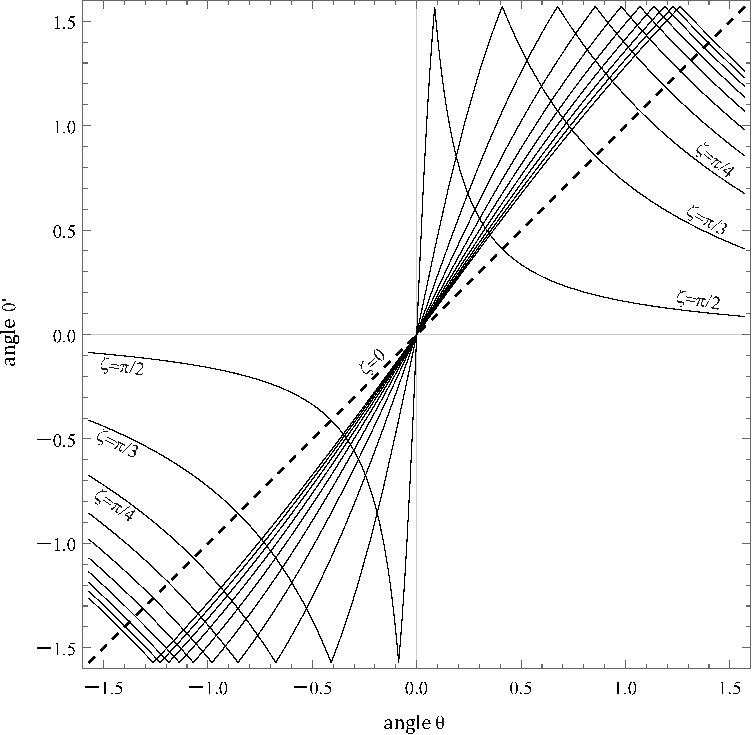}\caption{The alteration of the inclination angle in a frame moving away from
the source with velocity $\varsigma$. The dashed-line is the case
with the velocity $\varsigma=0$, hence $\theta'=\theta$. As the
velocity increases, so does the difference between $\theta'$ and
$\theta$. For a specific $\theta$ and $\varsigma$, a threshold
exists at $\theta'=\pm\frac{\pi}{2}$. For $\left|\theta'\right|\protect\leq\left|\pm\frac{\pi}{2}\right|$,
the angle diverge, represented by the slope $\protect\geq1$. However,
there exist combination values of $\theta$ and $\varsigma$ that
gives $\left|\theta'\right|\protect\geq\left|\pm\frac{\pi}{2}\right|$,
which implies that the angle spread ``backward'' in the opposite
direction of the observer's velocity $\varsigma$. Thus the observer
will perceive the angle as converging.}
\end{figure}
 It is worth noting that in a moving inertial frame, $\theta'$ is
time-independent, resulting in a constant angle deviation. 

For the case of an accelerated frame $\mathcal{A}$, the inclination
angle $\theta$ will be perceived as time-dependent, satisfying (\ref{eq:ab}).
Therefore, a similar beaming phenomenon as in the moving inertial
cases occurs. In the scenario where the accelerated observer approaching
the source, $\bar{\theta}$ decreases in time, resulting in a convergence
at $\tau\rightarrow\infty$, see FIG. 10. 
\begin{figure}
\centering{}\includegraphics[scale=0.65]{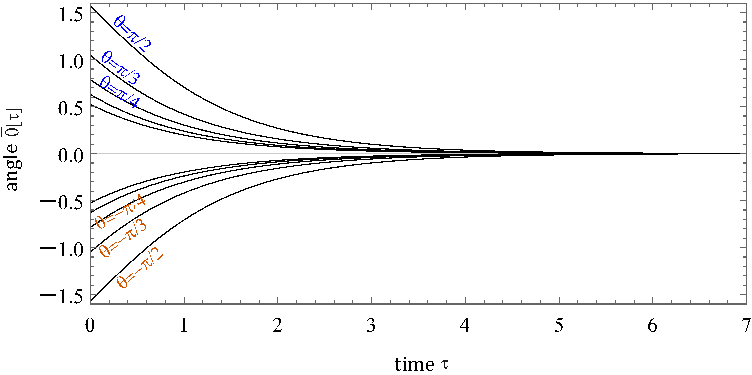}\caption{The evolution of the inclination angle perceived by an approaching
accelerated observer $\mathcal{A}$, namely $\bar{\theta}$, for different
value of $\theta$. $\bar{\theta}\rightarrow0$ as $\tau\rightarrow\infty$.}
\end{figure}
 In the scenario where the accelerated observer receding from the
source, $\theta$ will diverge until it reaches the maximal values,
namely $\pi/2$ for $0\leq\theta\leq\pi/2$ and $-\pi/2$ for $-\pi/2\leq\theta\leq0$.
From this point, it will stop diverging and start to converge as time
increases, as illustrated in FIG. 11. 
\begin{figure}
\centering{}\includegraphics[scale=0.65]{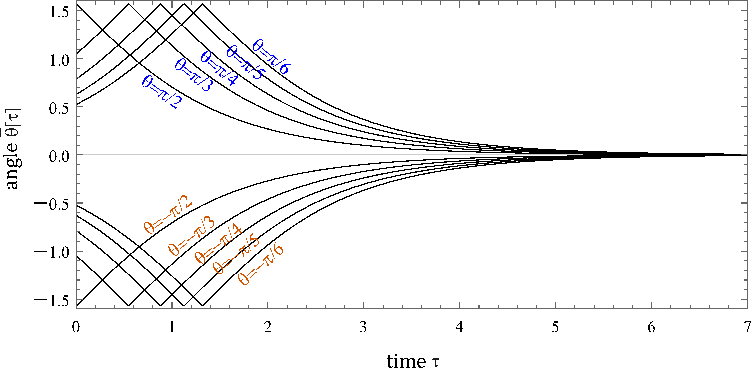}\caption{The evolution of the observation angle $\bar{\theta}$ with respect
to proper time $\tau$, as observed by an accelerated frame $\mathcal{A}$,
moving away from the source. The light rays initially dispersed at
some time interval but subsequently stopped diverging and started
to converge as time increased. Different initial angle $\theta$ will
results in different peaking-time, but the light rays will eventually
converge at $\tau\rightarrow\infty$.}
\end{figure}
 It is important to emphasize that our analysis of the relativistic
beaming phenomenon is restricted to specific cases, namely Cases 1
(accelerated away) and 4 (accelerated toward). The decelerated cases
(namely, Case 2 and 3) can be derived by inserting the condition (\ref{eq:awaytoward})
into (\ref{eq:ab}). In the context of relativistic beaming, Case
3 (decelerating toward) will experience a similar phenomenon to Case
1 (accelerating away), while Case 2 (decelerating away) is similar
to Case 4 (accelerating towards). Observing only the beaming of lightrays,
these two pairs of cases are indistinguishable. 

To be more precise, one could obtain the rate of change of the inclination
angle by inserting (\ref{eq:rateofchangeangle}) to (\ref{eq:ab})-
as follows:
\begin{equation}
\frac{d\bar{\theta}}{d\tau}=\frac{\sin\theta}{\cosh\bar{\varsigma}-\cos\theta\sinh\bar{\varsigma}}.\label{eq:itunglagi}
\end{equation}
Note that the inclination angle satisfies $-\frac{\pi}{2}\leq\theta\leq\frac{\pi}{2}$;
in these ranges, $\cosh\bar{\varsigma}$, $\cos\theta\sinh\bar{\varsigma}$
are positive definite, and the quantity $\left(\cosh\bar{\varsigma}-\cos\theta\sinh\bar{\varsigma}\right)^{-1}$
could be positive or negative, depending on whether $\cosh\bar{\varsigma}>\cos\theta\sinh\bar{\varsigma},$
or vice versa. For the case where the inclination angle at the rest
frame satisfies $0\leq\theta\leq\frac{\pi}{2}$, $\sin\theta$ will
be positive definite, so if the quantity $\left(\cosh\bar{\varsigma}-\cos\theta\sinh\bar{\varsigma}\right)^{-1}$
is positive, then $\frac{d\bar{\theta}}{d\tau}\geq0$; the angle at
$\mathcal{A}$ will diverge. Otherwise, $\frac{d\bar{\theta}}{d\tau}\leq0$,
and the angle will converge. Meanwhile, for the case of $-\frac{\pi}{2}\leq\theta\leq0$,
$\sin\theta$ will be negative definite, so the angle at $\mathcal{A}$
will converge if $\left(\cosh\bar{\varsigma}-\cos\theta\sinh\bar{\varsigma}\right)^{-1}$
is positive -resulting in $\frac{d\bar{\theta}}{d\tau}\leq0$-, otherwise
it will diverge.

\subsection{The Redshift and Blueshift}

Considering the quantity $\left\langle \mathbf{p},\mathbf{x}\right\rangle =p_{\mu}x^{\mu}$
with $x^{\mu}$ is a 4-vector and $p_{\mu}$ is a 4-momentum for a
single photon, one could show that $\left\langle \mathbf{p},\mathbf{x}\right\rangle $
is invariant under Lorentz transformation. A simple way to prove this
is to calculate $p_{\mu}x^{\mu}$ at frame $\mathcal{O}$ using (\ref{eq:photon1}),
and $p'_{\mu}x'^{\mu}$ at frame $\mathcal{O}'$ using (\ref{eq:photon2}).
With the help of (\ref{eq:Lorentz}), (\ref{eq:aberration-1})-(\ref{eq:aberration-2}),
and (\ref{eq:doppler2-2}), one could show that $p_{\mu}x^{\mu}=p'_{\mu}x'^{\mu}$.

Furthermore, with the invariance of $\left\langle \mathbf{p},\mathbf{x}\right\rangle $
in inertial frames, one can conclude that a plane wave moving in $\hat{n}$-direction
as follows:
\begin{equation}
\psi\left[x,t\right]=e^{\frac{i}{\hbar}\left\langle \mathbf{p},\mathbf{x}\right\rangle }=e^{i\omega\left(-t+x\right)},\label{eq:planewavefoton}
\end{equation}
is invariant under Lorentz transformation. Hence, a plane wave in
$\mathcal{O}$ is also a plane wave in $\mathcal{O}'$ but with the
change in frequency satisfying the Doppler effect (\ref{eq:doppler2-2}).
The Doppler effect causes the frequency in the inertial moving frame
$\mathcal{O}'$ to undergo a redshift if it is moving away from, -and
blueshift if it is approaching, the source. In the inertial frames,
the redshift and blueshift are constant in time.

However, this is not the case in an accelerated frame. Let us take
only the temporal part of (\ref{eq:planewavefoton}) at $\mathcal{O}$
for simplicity, namely $e^{-i\Theta},$ with $\Theta=\omega t$ is
the phase-angle. The accelerated observer $\mathcal{A}$ will measure
the photon of constant frequency $\omega$ at $\mathcal{O}$ as having
frequency $\bar{\omega}\left[\tau\right]$ satisfying the relativistic
Doppler effect (\ref{eq:doppler2}). Without loosing of generality,
let us set $\theta=0$ so that $\bar{\omega}\left[\tau\right]$ satisfies:
\begin{equation}
\bar{\omega}\left[\tau\right]=\omega e^{-\alpha\left(\xi\pm\tau\right)-\varsigma},\label{eq:special}
\end{equation}
considering only the case of $\bar{\varsigma}=\varsigma\pm\alpha\tau$
from (\ref{eq:awaytoward}). The phase-angle of the wave of the single
photon with respect to $\mathcal{A}$ is:
\begin{equation}
\bar{\Theta}\left[\tau\right]=\int\bar{\omega}\,d\tau=\mp\frac{\omega}{\alpha}e^{-\alpha\left(\xi\pm\tau\right)-\varsigma}=\mp\frac{\bar{\omega}\left[\tau\right]}{\alpha}.\label{eq:phaseaccel}
\end{equation}
$\bar{\Theta}$ is time-dependent, since $\bar{\omega}$ is a function
of time $\tau$. This is different with the phase-angle in the inertial
case, where $\Theta$ could be written simply as $\omega t$, since
$\omega$ in the inertial case is independent of time. Therefore,
a single mode of (the temporal part of) the radiation wave with respect
to $\mathcal{A}$ is:
\begin{equation}
\bar{\psi}\left[\tau,\xi\right]=e^{-i\bar{\Theta}}=e^{\pm i\frac{\omega}{\alpha}e^{-\alpha\left(\xi\pm\tau\right)-\varsigma}}.\label{eq:wavefunct}
\end{equation}
(\ref{eq:wavefunct}) is not necessarily a plane-wave, in contrast
to its inertial counterpart (\ref{eq:planewavefoton}), since the
frequency measured by $\mathcal{A}$ experiences a shift increasing
in time. The direction of the frequency's shift depends on the rate
of change of the frequency with respect to the proper time, namely:
\begin{equation}
\frac{d\bar{\omega}}{d\tau}=\mp\alpha\omega e^{-\alpha\left(\xi\pm\tau\right)-\varsigma};\label{eq:ratef}
\end{equation}
$\frac{d\bar{\omega}}{d\tau}<0$ gives a \textit{redshift} while $\frac{d\bar{\omega}}{d\tau}>0$
gives a \textit{blueshift}.

Now, having the definition of redshift and blueshift, let us consider
the 4 cases of accelerated observers in Section III A. Note that by
the relativistic Doppler effect, all these 4 cases will experienced
time-dependent shifts on their frequencies: the first and the third
cases will have their frequency shifting in time towards the infrared
direction (redshift), while the second and the last cases will have
the shift toward the ultraviolet direction (blueshift). These pairs
of cases could be distinguished by their instantaneous frequency with
respect to the frequency of the source at rest: the first case has
$\bar{\omega}\left[\tau\right]\leq\omega$, while the third has $\bar{\omega}\left[\tau\right]\geq\omega$,
although they experience increasing time-dependent redshift on their
frequencies. A similar condition occurs for the second case, with
$\bar{\omega}\left[\tau\right]\leq\omega$, and the last case, with
$\bar{\omega}\left[\tau\right]\geq\omega$; where both experience
increasing blueshift on their frequencies. Comparing the definition
on frequency's shift with the definition of acceleration and deceleration
in Section IIIA, it is clear that that they are not correlated, i.e.,
acceleration does not always correspond to redshift, while deceleration
does not always correspond to blueshift.

The \textquotedbl maximal\textquotedbl{} frequency $\omega_{\mathrm{max}}$
similarly transformed as (\ref{eq:special}), therefore the rate of
change of $\bar{\omega}_{\mathrm{max}}$ satisfies (\ref{eq:ratef}).
The peak of the spectrum will shift towards higher or lower frequencies
as time progresses, depending on which one from the 4 cases is satisfied.
See FIG. 12. 
\begin{figure}
\centering{}\includegraphics[scale=0.52]{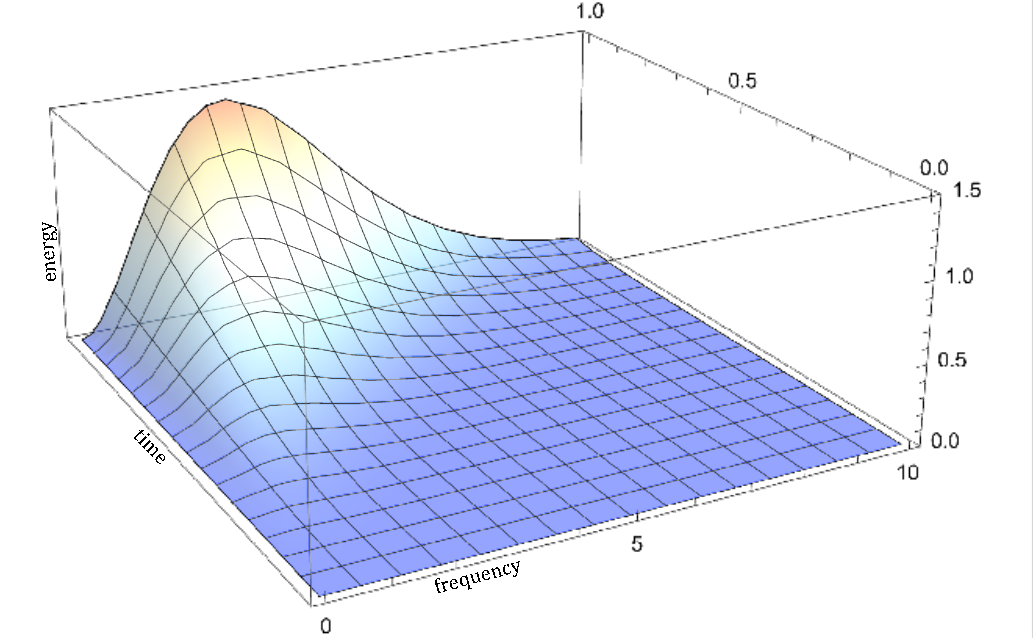}\caption{The 3D diagram of the ''Planck'' spectrum evolving in time, as perceived
by an accelerated observer $\mathcal{A}$, approaching the source
(Case 4). The $\left(x,y,z\right)$-axes are, respectively, the time
$\tau$, frequency $\bar{\omega}$, and energy density $\bar{\rho}{}_{E}$.
Notice that the dynamics in (\ref{eq:planckaccel-1}) is only observed
in the parameter of \textquotedbl maximal\textquotedbl{} frequency
$\bar{\omega}_{\mathrm{max}}.$}
\end{figure}

One may wonder the relationship between the frequency shift discussed
in this work and the gravitational red/blueshift phenomena \citep{Eddington}.
Gravitational frequency shift is an alteration in the photon's frequency
as it travels a gravitational field. The physical clock at different
location along the field has different rate of time, therefore, photon
appears to be red/blue-shifted relative to the frequency of the clock
\citep{Okun}. On the other hand, the frequency shift considered in
this work is due to its velocities; it has the same origin with the
relativistic Doppler effect, with the main difference lies in the
time-dependent nature of this shift. Now, by the equivalence principle,
that asserts the indistinguishability of gravity with acceleration,
one might expect an equivalency between the frequency shift due to
acceleration and gravitational red/blueshift. However, given a stationary
source of gravity, the gravitational redshift remains constant over
time, while the frequency shift due to acceleration is time-dependent.
It is worth noting that the equivalence principle holds only locally
in space \textit{and} in time, therefore if one considers a gravitational
field in a small region in space, it is approximately indistinguishable
with a uniform acceleration. This is also valid in reverse: if one
consider an acceleration in a (nearly) instantaneous time, it is indistinguishable
from a gravitational field. In this limit, the frequency shift due
to acceleration is practically indistinguishable to the gravitational
red/blueshift.

\subsection{Annihilation and Creation of Modes}

The number of modes/quanta of inertial moving frame, in general, will
differ with the number in a rest frame. This is due to the fact that
different inertial coordinates will measure different size of phase-space
volume element, hence, they will calculate different number of worldlines
that crosses the phase-space volume element. However, there will be
no creation or annihilation of quanta/modes, if their number $N$
is conserved in time. In contrast, the number of states $d\bar{N}$
in the accelerated frame evolves in time. One could obtain the rate
of the quanta/modes production as follows. Let us consider only the
accelerated and decelerated away case where $\bar{\varsigma}=\varsigma\pm\alpha\tau$.
Inserting (\ref{eq:doppler1}) to (\ref{eq:numberacc}), and setting
$\theta=0$ for simplicity gives: 
\[
\frac{d\bar{N}}{dN}=\frac{e^{-2\alpha\xi}}{2}\left(1+e^{-2\left(\varsigma\pm\alpha\tau\right)}\right):=\sigma\left[\alpha,\xi,\tau\right].
\]
From here we can obtain the rate of quanta/modes production in $\mathcal{A}$,
with respect to $\tau$:
\begin{equation}
\frac{d\sigma}{d\tau}=\mp\alpha e^{-2\alpha\xi}e^{-2\left(\varsigma\pm\alpha\tau\right)}=\mp\alpha\left(\frac{\bar{\omega}}{\omega}\right)^{2}.\label{eq:rate}
\end{equation}
For the case where $\mathcal{A}$ is accelerated away from the black-body
source at rest, the number of modes/quanta observed by $\mathcal{A}$
is decreasing exponentially in time, as described by the minus sign
in (\ref{eq:rate}). This corresponds to the increasing redshift experienced
by $\mathcal{A}$.  For the case where $\mathcal{A}$ is decelerated
away, the number of modes increases exponentially: Note that for $\frac{d\sigma}{d\tau}<0$,
the modes will be annihilated, while for $\frac{d\sigma}{d\tau}>0$,
the modes will be created in time.

\subsection{The Emissivity Factor}

Let us return to the Planck's law at an accelerated frame (\ref{eq:planckaccel}).
The scale factor $\varepsilon\left[\alpha,\xi\right]$ could be interpreted
physically as the emissivity resulted from the acceleration of the
observer. The emissivity is the factor that describes the imperfectness
of a black-body, i.e., a physical property that describes how efficiently
an object emits thermal radiation. The classical range of the emissivity
is $0\leq\varepsilon\leq1$, where 1 is the emissivity of a black-body
and 0 is the emissivity of a perfect thermal mirror, where no radiation
is emitted. In this interval, are the region of the grey-body. Classically,
$\varepsilon>1$ and $\varepsilon<0$ are not defined, except for
special cases for particles smaller than the dominant radiation wavelength
\citep{Golyk}.

There are two factors that affect the emissivity due to acceleration
as in (\ref{eq:emmissivity}), namely, the spatial coordinate $\xi$,
and the magnitude of the acceleration $\alpha$. Let us consider the
case where $\bar{\varsigma}=\varsigma\pm\alpha\tau$, namely the case
where the observer are respectively, accelerated and decelerated away
from the source. One could derive that for these cases, the scale
factor is:
\begin{equation}
\varepsilon\left[\alpha,\xi\right]=e^{2\alpha\xi}\left(1\mp\alpha\right),\label{eq:emmissivity-1}
\end{equation}
where the minus sign is for the accelerated away case, and the plus
for the decelerated. For the accelerated away case, the region with
negative emissivity is obtained when $\alpha>1$; at this condition,
no radiation will be detected by the accelerated observer. Meanwhile,
positive emissivity factor is obtained when $0\leq\alpha\leq1$. For
this case, if $\xi>0,$ it is possible to have the emissivity $\varepsilon>1$,
hence the black-body will be perceived as ``hyperblack'' in the
accelerated frame; otherwise, for $\xi<0$, it will be perceived as
grey. See FIG. 13. 
\begin{figure}
\begin{centering}
\includegraphics[scale=0.65]{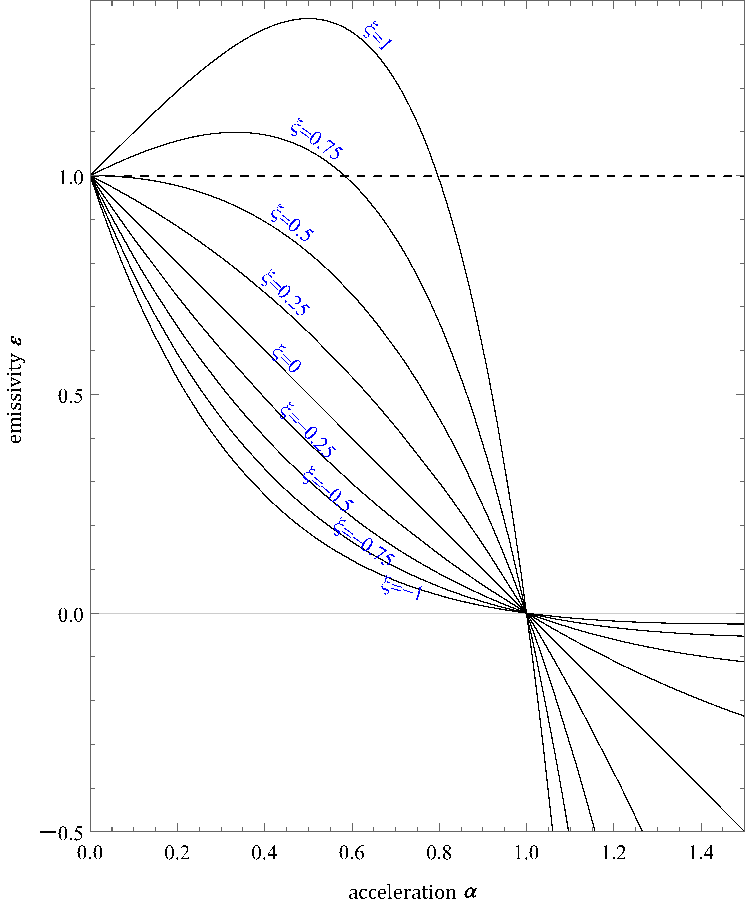}
\par\end{centering}
\caption{The plot of the source's emissivity $\varepsilon$ with respect to
the acceleration magnitude $\alpha$, for different value of spatial
position $\xi$. There is a strict threshold $\alpha\protect\geq1$
where there is no radiation detected by the accelerated observer due
to $\varepsilon\protect\leq0$, independent from the value of $\xi$.
For $0\protect\leq\alpha\protect\leq1$, the source could be perceived
as grey, black, or hyperblack, depending on $\xi$. The dashed-line
is the condition of the black-body with $\varepsilon=1$; above and
below the line are, respectively, the 'hyperblack' and grey regions.
For $\xi=0$, the relation between emissivity and acceleration becomes
linear: $\varepsilon=1-\alpha$.}
\end{figure}
 For the decelerated away case, the emissivity factor $\varepsilon\geq0$,
so the accelerated frame will observe a range of hyperblack, black,
and greybody, depending on the value of $\alpha$ and $\xi$. See
FIG. 14. 
\begin{figure}
\begin{centering}
\includegraphics[scale=0.65]{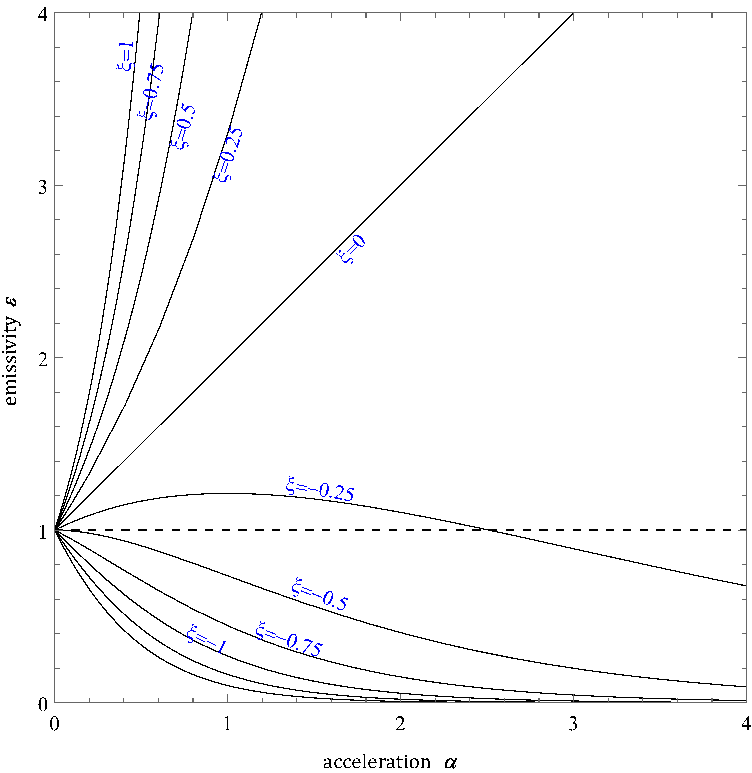}
\par\end{centering}
\caption{The plot of the source's emissivity $\varepsilon$ with respect to
the \textit{deceleration} magnitude $\alpha$, for different value
of spatial position $\xi$. The emissivity for this case is positive
definite: $\varepsilon\protect\geq0$, hence the observer will detect
radiation for any value of $\alpha$. The source could be perceived
as grey, black, or hyperblack, depending on $\xi$. The dashed-line
is the condition of the black-body with $\varepsilon=1$, above and
below the line are, respectively, the 'hyperblack' and grey region.
For $\xi=0$, the relation between emissivity and acceleration becomes
linear: $\varepsilon=1+\alpha$.}
\end{figure}

While in the decelerated away case one can perceive the source with
different 'blackness' depending on the value of $\alpha$ and $\xi$,
an interesting phenomenon occurs in the case of an accelerated away
scenario. For this case, if $0\leq\alpha\leq1$ is satisfied, the
emissivity is $0\leq\varepsilon\leq1$; the black-body is perceived
as a greybody in $\mathcal{A}$, where the signal becomes dimmer as
$\alpha\rightarrow1$. At $\alpha=1$, $\varepsilon=0$: in the frame
of $\mathcal{A}$, the black-body stops emitting radiation as if it
is a perfect thermal mirror. From this point, the emissivity can only
be negative definite as $\alpha$ increases. Remarkably, for a constant
$\xi$, the emissivity is independent of time and the position of
the observer from the source, which means that no matter how far or
close the object is from the observer, theoretically, they will measure
the same emissivity. There could be several possible explanations
for this phenomenon, which might include the relativistic beaming
and the existence of the (Killing) horizon at $x=\pm t$, however,
further research is needed to understand this phenomenon completely.

\subsection{Wien's Displacement Law and Temperature}

At the moment, we are not able to derive the transformation of Wien's
displacement law in a moving frame without knowing how the temperature
$T$ in relation (\ref{eq:WienLaw}) transform under Lorentz transformation.
However, if we assume that Wien's law is valid also in a moving frame,
we can define the directional temperature using the Wien's law (\ref{eq:WienLaw}),
simply as a quantity proportional to the frequency that maximies the
energy of the source. This, at least, will be useful for the calculation
of a temperature for a black-body in moving frames. As we had mentioned
in the Introduction, any measurement of temperature in moving bodies
are measured via it's radiation, since we are not able to do a direct
measurement of the temperature in moving bodies. If a body is moving,
it needs to be at rest with respect to an observer so that it could
be thermalize/ in a thermal equilibrium with the measurement apparatus.
Here, satisfied or not, the temperature for moving bodies, is treated
as a derived variables. Moreover, if we make an assumption that (\ref{eq:WienLaw})
is valid for inertial and moreover, uniformly accelerated frame, then
the directional temperature $T$ must transform in the same way as
the frequency, namely:
\begin{equation}
\bar{T}=\frac{Te^{-\alpha\xi}}{\cosh\bar{\varsigma}\left[\tau\right]+\cos\bar{\theta}\left[\tau\right]\sinh\bar{\varsigma}\left[\tau\right]}.\label{eq:directempt}
\end{equation}
(\ref{eq:directempt}) is a generalization of the 'directional'-(effective)
temperature in \citep{Henry} which is valid in any inertial frame;
it is an extension of the result in \citep{Henry} to a uniformly
accelerated frame. One could observe that in (\ref{eq:directempt}),
there is a coordinate-dependent scale factor proportional to $\sim e^{-\alpha\xi},$
similar to the temperature scale factor in Tolman's theory of general
relativistic thermodynamics \citep{Tolman}.

\section{Discussion and Conclusion}

\subsection{Discussion}

The Planck's formula in the accelerated frame (\ref{eq:planckaccel})
are based on several assumptions in the derivation, hence, if one
or more of these assumptions are not valid, this will also affect
the validity of (\ref{eq:planckaccel}). Let us list the assumptions
we use for the derivation:
\begin{enumerate}
\item $E=n\hbar\omega$. This is implicitly used because we started from
the Planck's law in rest frame (the relation is use to derive Planck's
law).
\item The validity of the de Broglie postulate $p=\hbar k$ in the accelerated
frame, namely: $p\left[\tau,\xi\right]=\hbar k\left[\tau,\xi\right]$,
for every Rindler time $\tau$ and Rindler position $\xi$.
\item The invariance of the relativistic distribution function (number of
world-lines that crosses a hypersurface) under coordinate transformation.
\end{enumerate}
Starting from these assumptions, together with some standard definitions
in the theory of relativity, all the equations we derived, most importantly,
the relativistic Doppler effect (\ref{eq:doppler2}), the transformation
of solid angle (\ref{eq:solidangleacc}), the volume contraction (\ref{eq:volumecontractionrindler}),
and the transformation of the number of modes distribution (\ref{eq:numberacc}),
are consequences of the assumptions and definitions we used. Let us
check if all these assumptions are reasonable. The first assumption
is used to derive Planck's law in the original form, and since it
is not use explicitly to derive the results, this should not concern
our work. The second assumption is used explicitly to derive the dispersion
relation (\ref{eq:disperssionrel}) from the momentum constraint $p_{\mu}p^{\mu}=0.$
(\ref{eq:disperssionrel}) is crucial for deriving the momentum in
$\mathcal{A},$ namely (\ref{eq:momentumrindler}). However, one can
also obtained the dispersion relation (\ref{eq:disperssionrel}) without
the de Broglie postulate, if we consider the electromagnetic wave
equation $\boxempty\mathbf{E}=0$, with $\boxempty$ is the d'Alembertian
operator. The form of the wave equation and its solution are invariant
under Rindler coordinate transformation, and so does the dispersion
relation. Therefore, without the de Broglie postulate, (\ref{eq:disperssionrel})
is still valid in $\mathcal{A}$ as long as the Maxwell equation is
satisfied by the black-body radiation. 

Finally, opinions on the last assumption are divided among the researchers
in the field. The relativistic distribution function $f\left(\mathbf{x},\mathbf{p}\right)$
is defined as (\ref{eq:distrirel}). Since our work follows closely
the work of \citep{Peebles,Heer,Henry}, we use the same assumption
that the relativistic distribution function $f\left(\mathbf{x},\mathbf{p}\right)$
is coordinate-invariant, instead of the invariance of the number of
states $dN$, which is also employed in many existing works in the
literature. The reason for this difference is because there are 2
distinct measurements of volumes in moving frames: the ones that measure
all the points in the volume simultaneously at an instant time, and
the one that measure all the points in the volume at a time interval
$\Delta\tau=\tau_{f}-\tau_{i}$. Our work use the second type of measurement,
and hence the number of states counted by the 'moving volume' at time
interval $\Delta\tau$ is not the same as $dN$ (the number of states
in rest frame); hence we use assumption 3.

Some readers may wonder if a simple coordinate transformation from
inertial to Rindler coordinate is sufficient to obtain the physics
in an accelerated frame. As far as we understand, in the theory of
relativity (special or general), the use of different coordinates
on spacetime, in general, will result in different observational perspectives.
This is not only valid for coordinates related by Lorentz transformation,
but also for more general coordinate systems, including the Rindler
coordinate. Therefore, to obtain observational perspective from an
accelerated frame, it is sufficient to do a transformation from inertial/Cartesian
coordinate to Rindler coordinate. Furthermore, one might also wonder
why the Unruh effect is not included in our work. This is out of the
scope of the subject in our work, since the Unruh effect is derived
from the quantum field theoretic derivation, and we have tried to
write our paper as classical as possible.\textbf{ }The reason to write
the paper as classical as possible is because the subject is full
of controversy, so in our opinion, it will be an advantage if we could,
at least, clearly understand the subject in the (semi)-classical level.
Although it might be possible to include the Unruh effect in the discussion,
in the classical level, it can not be obtained just by coordinate
transformation from Minkowski/Cartesian coordinate to Rindler coordinate.
There is another assumption needed to obtain the Unruh effect in the
classical setting, and this is related to the definition of the classical
vacuum. This will be an insteresting subject to pursue.

\subsection{Conclusion}

Let us conclude our work in this article as follows. We have derived
Planck's law and calculated the black-body spectrum in a uniformly
accelerated frame using Rindler coordinates. The spectrum is time-dependent,
Planckian at each instantaneous time, but it is scaled by an emissivity
factor $\varepsilon=e^{\alpha\xi}\left(1\pm\alpha\right)$ that depends
on the Rindler spatial coordinate $\xi$ and the acceleration magnitude
$\alpha$. The spatial, coordinate-dependent scale factor is proportional
to $\sim e^{\alpha\xi},$ while the scale factor related to the acceleration
is proportional to $\sim e^{\alpha\xi}\left(1\pm\alpha\right)$, depending
on the observer is either accelerated or decelerated. An observer
decelerating away from the source will perceive the black-body as
hyperblack, black, or grey, while for the accelerating-away observer,
there is a limit in the acceleration magnitude in receiving the radiation,
namely, if $\alpha\geq1$, the accelerated observer will stop receiving
radiation from the black-body. Outside of this limit, the black-body
is perceived either as grey or hyperblack, depending on the spatial
Rindler coordinate $\xi$. 

The variables in the Planck's law, specifically the number of modes
and frequency, evolve over time. The Planckian spectrum is continuously
red-shifted towards lower frequencies as time progresses for the case
where the observer is accelerating-away or decelerating towards, and
blue-shifted towards lower (or higher) frequencies for the case where
the observer is accelerating-towards or decelerating-away. In the
accelerated frame, the production of radiation modes (photons) can
be positive or negative, depending on the acceleration or deceleration
of the observer, and zero for vanishing acceleration. Besides this,
there exists a peculiar phenomenon perceived by the accelerating observer,
the relativistic beaming, where the rays of light tends to converge,
or diverge in the direction of motion, depending on whether the observer
is moving towards, or away from the source. In the end, assuming Wien's
displacement law also holds in the accelerated frame, the time-dependent,
(directional)-temperature of a body within an accelerated frame is
given by $\bar{T}\left[\tau\right]=e^{-\alpha\xi}\left(\cosh\bar{\varsigma}\left[\tau\right]+\cos\bar{\theta}\left[\tau\right]\sinh\bar{\varsigma}\left[\tau\right]\right)^{-1}T.$

Our last comment is on the temperatures of the systems. For the moment,
we do not have a universal definition of temperature. However, the
(effective) temperature (\ref{eq:directempt}) is derived from their
frequency spectra, this is also the case for the Hawking-Unruh temperature.
Since for a moving body we do not have other choice than treating
(effective) temperature as a derived quantity, we argue that one needs
to take seriously the directional approach of temperature. This is
supported by the argument in \citep{Aldrovandi} that the inertial
version of effective temperature (\ref{eq:directempt}) is not merely
a mathematical parameter, but a real transformation law. We are optimistic
that this will be a useful approach to understanding the nature of
temperature.

\begin{acknowledgments}

S. A. was supported by an appointment to the Young Scientist Training
Program at the Asia Pacific Center for Theoretical Physics (APCTP)
through the Science and Technology Promotion Fund and Lottery Fund
of the Korean Government. This was also supported by the Korean Local
Governments - Gyeongsangbuk-do Province and Pohang City. H. L. P.
would like to thank Ganesha Talent Assistanship (GTA) Institut Teknologi
Bandung for financial support.

\end{acknowledgments}

\appendix

\section{Black-Body Radiation in Rest Frame}

\subsection{Planck's Radiation Law}
To make this article self-contained, in this subsection of the Appendix
A, we derive the Planck's law in the form of relation (\ref{eq:Plancklawnew})
from the original equation \citep{Planck2}:
\begin{equation}
B\left[f,T\right]=\frac{2hf^{3}}{c^{2}}\frac{1}{e^{\nicefrac{hf}{\left(k_{B}T\right)}}-1}.\label{eq:Plancklaw}
\end{equation}
$B$ is the \textit{intensity} or \textit{spectral radiance}, defined
as:
\begin{equation}
B\left[f\right]=\frac{dE}{\cos\theta dfdAd\Omega dt},\label{eq:intensity}
\end{equation}
with $f$ is the frequency of the electromagnetic radiation and $T$
is the temperature of the black-body source emitting the radiation.
$h,$ $c$, and $k_{B}$ are respectively, the Planck constant, the
speed of light, and the Boltzmann constant. The intensity $B\left[f\right]$
is defined as the infinitesimal energy $dE$ that passes through an
infinitesimal area $dA\,d\Omega$ of the surface of a sphere, per
time interval $dt$, in the frequency range $\left[f,f+df\right]$.
$\theta$ is the \textit{inclination angle}: the angle between the
velocity of the photon (that is parallel to the normal to $dA$) with
the direction of the observation, but with frame $\mathcal{A}$ replaced
with frame $\mathcal{O}'$. Inserting (\ref{eq:intensity}) to (\ref{eq:Plancklaw}),
and multiplying the right hand side of the equation with $\frac{c}{c}=1$
gives:
\begin{equation}
dE=\frac{2hf^{3}}{c^{3}}\frac{1}{e^{\nicefrac{hf}{\left(k_{B}T\right)}}-1}dfd\Omega dA\,c\cos\theta dt.\label{eq:Plancklaw3}
\end{equation}
The quantity $dA\,c\cos\theta dt$ is the infinitesimal volume $dV$
swept out by the radiation. Using the angular frequency $\omega=2\pi f$
instead of $f$, the Planck's law can be written as:
\begin{equation}
dE=\underset{\rho_{E}\left[\omega\right]}{\underbrace{\frac{\hbar\omega^{3}}{4\pi^{3}c^{3}}\frac{1}{e^{\nicefrac{\hbar\omega}{\left(k_{B}T\right)}}-1}}}d\omega d\Omega dV.\label{eq:Plancklaw4}
\end{equation}
Moreover, given the energy of each frequency mode $\omega$ as $dE=\hbar\omega dN$,
with $N$ is the number (or distribution) of the radiation modes (quanta
of radiation/photon, however, we avoid such terminologies because
we want this paper to be as classical as possible) having an angular
frequency $\omega$, we could obtain the Planck's distribution:
\begin{equation}
dN=\frac{\omega^{2}}{4\pi^{3}c^{3}}\frac{1}{e^{\nicefrac{\hbar\omega}{\left(k_{B}T\right)}}-1}d\omega d\Omega dV,\label{eq:Plancklaw5}
\end{equation}
with the famous Planckian spectrum. (\ref{eq:Plancklaw5}) is the
form of Planck's law used in \citep{Henry}.

\subsubsection{Wien's Displacement Law}

To derive Wien's displacement law from the Planck's law (\ref{eq:Plancklaw4}),
one needs to find the angular frequency $\omega_{\mathrm{max}}$ that
maximize the radiation energy $E$.  This can be obtained by solving
the equation $\frac{d\rho_{E}\left[\omega\right]}{d\omega}=0$ as
follows:
\begin{equation}
\frac{\hbar}{4\pi^{3}c^{3}}\frac{\omega^{2}}{e^{\nicefrac{\hbar\omega}{\left(k_{B}T\right)}}-1}\left(3-\frac{\hbar\omega}{k_{B}T}\frac{e^{\nicefrac{\hbar\omega}{\left(k_{B}T\right)}}}{e^{\nicefrac{\hbar\omega}{\left(k_{B}T\right)}}-1}\right)=0.\label{eq:dPlanck}
\end{equation}
Let us define $w=\frac{\hbar\omega}{k_{B}T}$, the equation from (\ref{eq:dPlanck})
that we need to solve becomes:
\begin{equation}
w=3\left(1-e^{-w}\right),\label{eq:Lambert}
\end{equation}
which can be solved by the Lambert function $W_{0}$ as follows:
\begin{equation}
w=3+W_{0}\left(-3e^{-3}\right).\label{eq:solution}
\end{equation}
With this, one could obtain the Wien's displacement law in terms of
angular frequency:
\begin{equation}
k_{B}T=\frac{1}{\left(3+W_{0}\left(-3e^{-3}\right)\right)}\hbar\omega_{\mathrm{max}},\label{eq:WienLaw}
\end{equation}
with $\omega_{\mathrm{max}}$ is the angular frequency that maximize
the radiation energy $E$, the 'maximal' frequency. The statement
that the temperature of a black-body source is proportional to the
frequency that maximizes the energy of the radiation comes directly
from (\ref{eq:WienLaw}).

One can insert the Wien's law (\ref{eq:WienLaw}) to (\ref{eq:Plancklaw5})
and obtain:
\begin{equation}
dN=\frac{\omega^{2}}{4\pi^{3}c^{3}}\frac{1}{e^{\nicefrac{w\omega}{\omega_{\mathrm{max}}}}-1}d\omega d\Omega dV.\label{eq:Plancklawnew-2}
\end{equation}
We called the dimensionless coefficient $w$ satisfying (\ref{eq:Lambert})
as the Wien constant. Notice that for a different representation of
the Planck's law, i.e., if the law written as a function of the wavelength
$\lambda$ instead of $\omega$, the Wien 'constant' will differ.

The Planck's law in the form of equation (\ref{eq:Plancklaw5}) is
equivalent with the two equations: the ``Planck's law'' in terms
of maximal frequency (\ref{eq:Plancklawnew-2}), together with the
Wien's law (\ref{eq:WienLaw}). To avoid the problem of the temperature
in moving bodies \citep{Farias,Derakshani}, we will use the ``Planck's
law'' (\ref{eq:Plancklawnew-2}) in our analysis for the black-body
spectrum in a uniformly accelerated frame.

\section{Black-Body Radiation in Inertial Frames}

\subsection{Lorentz Transformation}

In this subsection, we review the basic properties of Lorentz transformation
and define the notations used in our article. Let $\mathcal{O}$ be
a frame at rest, with spatial coordinates $\left(x,y,z\right)$ and
time coordinate $\mathfrak{t}.$ In the covariant approach, the time
and space are regarded in an equal manner, so let us define the 'time'
$t:=c\mathfrak{t}$ such that $t$ has the same dimension as length;
$c$ is the speed of light. Let $\mathbf{x}$ =$\left(x,y,z,t\right)$
be an inertial coordinate that parametrizes the Minkowski space.

Suppose another inertial frame $\mathcal{O}'$ is moving with a constant
velocity $u$ in the $x$-direction. Let $\mathbf{x}'=\left(x',y',z',t'\right)$
be another inertial coordinate of the Minkowski space related to frame
$\mathcal{O}'$. Frames $\mathcal{O}$ and $\mathcal{O}'$ are related
by the coordinate (Lorentz) transformation as follows:
\begin{equation}
\begin{array}{cc}
t' & =\gamma\left(t-\beta x\right),\\
x' & =\gamma\left(x-\beta t\right),
\end{array}\quad\begin{array}{cc}
t & =\gamma\left(t'+\beta x'\right),\\
x & =\gamma\left(x'+\beta t'\right),
\end{array}\label{eq:Lorentz}
\end{equation}
where $y,z$ are not affected by the transformation, namely $y'=y$
and $z'=z.$ Here, we use $\beta=\frac{u}{c}$ and $\gamma=\frac{1}{\sqrt{1-\beta^{2}}}.$
Let us define the \textit{rapidity} as $\varsigma=\cosh^{-1}\gamma$,
then we have the following relations:
\begin{equation}
\gamma=\cosh\varsigma,\;\beta\gamma=\sinh\varsigma,\;\beta=\tanh\varsigma.\label{eq:hyperbolicfunct}
\end{equation}
With relations (\ref{eq:hyperbolicfunct}), the Lorentz transformation
(\ref{eq:Lorentz}) can be written in terms of hyperbolic functions:
\begin{equation}
\begin{array}{cc}
t' & =t\cosh\varsigma-x\sinh\varsigma,\\
x' & =x\cosh\varsigma-t\sinh\varsigma,
\end{array}\;\begin{array}{cc}
t & =t'\cosh\varsigma+x'\sinh\varsigma,\\
x & =x'\cosh\varsigma+t'\sinh\varsigma.
\end{array}\label{eq:Lorentz-1}
\end{equation}

The infinitesimal (4-vector) transformation related
to (\ref{eq:Lorentz-1}) can be obtained as follows:
\begin{equation}
\begin{array}{ccc}
dt' & =\gamma\left(dt-\beta dx\right) & =dt\cosh\varsigma-dx\sinh\varsigma\\
dx' & =\gamma\left(dx-\beta dt\right) & =dx\cosh\varsigma-dt\sinh\varsigma
\end{array},\label{eq:Lorentz-2}
\end{equation}
together with their inverses:
\begin{equation}
\begin{array}{ccc}
dt & =\gamma\left(dt'+\beta dx'\right) & =dt'\cosh\varsigma+dx'\sinh\varsigma\\
dx & =\gamma\left(dx'+\beta dt'\right) & =dx'\cosh\varsigma+dt'\sinh\varsigma
\end{array},\label{eq:Lorentz-3}
\end{equation}
with $dy'=dy$ and $dz'=dz$.

\subsection{The Relativistic Aberration of Light}

Suppose with respect to the rest frame $\mathcal{O}$ we have an object
moving with velocity $\mathbf{v}$ as follows:
\[
\mathbf{v}=\left(v_{x},v_{y},v_{z}\right)=\left(\frac{dx}{dt},\frac{dy}{dt},\frac{dz}{dt}\right).
\]
In frame $\mathcal{O}'$, the velocity of the object is:
\[
\mathbf{v}'=\left(v'_{x},v'_{y},v'_{z}\right)=\left(\frac{dx'}{dt'},\frac{dy'}{dt'},\frac{dz'}{dt'}\right).
\]
Using the transformation (\ref{eq:Lorentz-2})-(\ref{eq:Lorentz-3}),
one could obtain the relation between $\mathbf{v}$ and $\mathbf{v}'$
as follows
\begin{equation}
v_{x}=\frac{v'_{x}+\beta}{1+\beta v'_{x}}=\frac{v'_{x}+\tanh\varsigma}{1+v'_{x}\tanh\varsigma},\label{eq:velocity1}
\end{equation}
\begin{equation}
v_{y,z}=\frac{v'_{y,z}}{\gamma\left(1+\beta v'_{x}\right)}=\frac{v'_{y,z}}{\cosh\varsigma\left(1+v'_{x}\tanh\varsigma\right)},\label{eq:velocity1yz}
\end{equation}
and their inverses:
\begin{equation}
v'_{x}=\frac{v_{x}-\beta}{1-\beta v{}_{x}}=\frac{v_{x}-\tanh\varsigma}{1-v_{x}\tanh\varsigma},\label{eq:velocity2}
\end{equation}
\begin{equation}
v'_{y,z}=\frac{v_{y,z}}{\gamma\left(1-\beta v{}_{x}\right)}=\frac{v_{y,z}}{\cosh\varsigma\left(1-v_{x}\tanh\varsigma\right)}.\label{eq:velocity2yz}
\end{equation}

Let a black-body source be  at rest with respect to an inertial frame
$\mathcal{O}.$ The black-body are emitting electromagnetic radiation
(photons) with propagation velocity $c$ in the direction of $\hat{n}=\left(\cos\theta,\sin\theta,0\right)$,
with $\theta$ is the (polar) inclination angle between the axis +$x$
and $\hat{n}$ on plane $dx\wedge dy$, but with frame $\mathcal{A}$
replaced with frame $\mathcal{O}'$. The electric and magnetic part
of the radiation lie, respectively, in plane $dx\wedge dy$ and $\hat{n}\wedge dz$,
so the propagation velocity will not affect their amplitude. The velocity
of the photon with respect to frame $\mathcal{O}$ is: 
\begin{equation}
\mathbf{v}=\left(v_{x},v_{y},v_{z}\right)=c\hat{n}=\left(c\cos\theta,c\sin\theta,0\right).\label{eq:velo1}
\end{equation}

Meanwhile, at frame $\mathcal{O}'$ , the velocity of the photon is:
\begin{equation}
\mathbf{v}'=\left(v'_{x},v'_{y},v'_{z}\right)=c\hat{n}'=\left(c\cos\theta',c\sin\theta',0\right),\label{eq:velo2}
\end{equation}
with $\hat{n}'$ and $\theta'$ are the propagation direction and
the inclination angle of the photon according to $\mathcal{O}'$,
respectively. This could be derived from the null (light-like) vector
condition for light, where the norm of its 4-velocity $\mathbf{u}=\left(\gamma,\gamma\mathbf{v}\right)$
is always zero in any coordinate system. This gives:
\[
u^{\alpha}u_{\alpha}=\gamma^{2}\left(\left\Vert \mathbf{v}\right\Vert ^{2}-1\right)=0,
\]
where the 3-velocities satisfy (\ref{eq:velo1}) and (\ref{eq:velo2}).

Using the velocity addition formulas in (\ref{eq:velocity1})-(\ref{eq:velocity2}),
one could obtain the transformation between velocities of the photon
seen by $\mathcal{O}$ and $\mathcal{O}'$, which results in the polar
angle formulas as follows:
\begin{align}
\cos\theta' & =\frac{\cos\theta-\beta}{1-\beta\cos\theta},\label{eq:aberration-1}\\
\sin\theta' & =\frac{\sin\theta}{\gamma\left(1-\beta\cos\theta\right)}.\label{eq:aberration-2}
\end{align}
Inserting (\ref{eq:aberration-1}) and (\ref{eq:aberration-2}) to
a trigonometric identity, $\tan\frac{\theta'}{2}=\frac{\sin\theta'}{1+\cos\theta'}$,
gives:
\begin{equation}
\tan\frac{\theta'}{2}=\frac{1}{\gamma\left(1-\beta\right)}\tan\frac{\theta}{2}=\frac{\sqrt{1+\beta}}{\sqrt{1-\beta}}\tan\frac{\theta}{2}.\label{eq:aberration-3}
\end{equation}
Equation (\ref{eq:aberration-1})-(\ref{eq:aberration-3}) are the
aberration of light formulas \citep{Johnson}. Their inverses are:
\begin{align}
\cos\theta & =\frac{\cos\theta'+\beta}{1+\beta\cos\theta},\label{eq:aberration-1-1}\\
\sin\theta & =\frac{\sin\theta'}{\gamma\left(1+\beta\cos\theta'\right)},\label{eq:aberration-2-1}\\
\tan\frac{\theta}{2} & =\frac{\sqrt{1-\beta}}{\sqrt{1+\beta}}\tan\frac{\theta'}{2}.\label{eq:aberration-3-1}
\end{align}

\subsection{The Relativistic Doppler Effect}

The next step is to derive the relativistic Doppler effect. First,
we need to define the 4-momentum $\mathbf{p}$ of a moving body with
respect to $\mathcal{O}$ as follows:
\begin{equation}
\mathbf{p}=\left(p_{t},\underset{^{3}\mathbf{p}}{\underbrace{p_{x},p_{y},p_{z}}}\right),\label{eq:momentum}
\end{equation}
with $p_{t}=\nicefrac{E}{c}$ is the energy of the moving body and
$^{3}\mathbf{p}$ is the relativistic 3-momentum. For the black-body
radiation case, the moving bodies are photons, which, by de Broglie
postulate, has energy $E=\hbar\omega$ and momentum $^{3}\mathbf{p}=\hbar\mathbf{k}$.
Here, $\hbar=\nicefrac{h}{2\pi}$ is the Planck constant, $\omega=2\pi f$
is the (angular) frequency, and $\mathbf{k}=\left(k_{x},k_{y},k_{z}\right)$
is the wave vector satisfying $\mathbf{k}=\frac{2\pi\hat{n}}{\lambda}.$
Inserting these information to (\ref{eq:momentum}) together with
the components of $\hat{n}$, one could obtain the 4-momentum of a
photon with respect to an inertial observer $\mathcal{O}$:
\begin{equation}
\mathbf{p}=\frac{\hbar\omega}{c}\left(1,\hat{n}\right)=\frac{\hbar\omega}{c}\left(1,\cos\theta,\sin\theta,0\right).\label{eq:photon1}
\end{equation}
According to the moving frame $\mathcal{O}'$, the 4-momentum of the
photon is:
\begin{equation}
\mathbf{p}'=\left(p_{t}',p_{x}',p_{y}',p_{z}'\right)=\frac{\hbar\omega'}{c}\left(1,\cos\theta',\sin\theta',0\right),\label{eq:photon2}
\end{equation}
with $\omega'$ is the frequency of the photon, with respect to $\mathcal{O}'$.
One could also obtain (\ref{eq:photon1}) and (\ref{eq:photon2})
by inserting the de Broglie postulate to the zero mass condition for
the photon:
\[
p^{\mu}p_{\mu}=-E^{2}+\left\Vert ^{3}\mathbf{p}\right\Vert ^{2}=0;
\]
with this, we only consider the positive solution from the dispersion
relation $\omega^{2}=c^{2}k^{2}.$

The 4-momentum is an element of the Minkowski space and therefore
transform under Lorentz transformation, so the relation of $\mathbf{p}'$
and $\mathbf{p}$ is:
\begin{equation}
\begin{array}{cc}
p_{t}' & =\gamma\left(p_{t}-\beta p_{x}\right),\\
p_{x}' & =\gamma\left(p_{x}-\beta p_{t}\right),
\end{array}\quad\begin{array}{cc}
p_{t} & =\gamma\left(p_{t}'+\beta p_{x}'\right),\\
p_{x} & =\gamma\left(p_{x}'+\beta p_{t}'\right),
\end{array}\label{eq:Lorentz-4}
\end{equation}
with $p_{y}'=p_{y}$ and $p_{z}'=p_{z}$. Naturally, the (generalized)
momentum is a co-vector instead of a (contravariant) vector, however,
here we use the vector transformation of momentum. This will not affect
the result as long as the calculations are consistent. Using the Lorentz
transformation (\ref{eq:Lorentz-4}) on the momenta (\ref{eq:photon1})
and (\ref{eq:photon2}), one could obtain the relations between the
frequency of photon $\omega$ as seen by $\mathcal{O}$ and $\omega'$
as seen by $\mathcal{O}'$, written in 3 equivalent forms as follows
\citep{Johnson}:
\begin{align}
\frac{\omega}{\omega'} & =\frac{\sin\theta'}{\sin\theta},\label{eq:doppler0}\\
\frac{\omega}{\omega'} & =\gamma\left(1+\beta\cos\theta'\right),\label{eq:doppler}\\
\frac{\omega}{\omega'} & =\frac{\gamma\left(\cos\theta'+\beta\right)}{\cos\theta}.\nonumber 
\end{align}
From these equations, one could retrieve the aberration formulas (\ref{eq:aberration-1})-(\ref{eq:aberration-3})
and their inverses (\ref{eq:aberration-1-1})-(\ref{eq:aberration-3-1}).
Inserting (\ref{eq:aberration-1}) to (\ref{eq:doppler}), one could
rewrite $\frac{\omega}{\omega'}$ in terms of only variable $\theta$,
the inclination angle seen by observer $\mathcal{O}$:
\begin{align}
\frac{\omega}{\omega'} & =\frac{1}{\gamma\left(1-\beta\cos\theta\right)}.\label{eq:doppler2-2}
\end{align}
Equation (\ref{eq:doppler}) and (\ref{eq:doppler2-2}) are the formula
describing the relativistic Doppler effect for the frequency of the
photon in a moving observer $\mathcal{O}'$ with respect to observer
at rest $\mathcal{O}$ \citep{Johnson}. For our case, the photon
frequency observed by the moving observer $\mathcal{O}'$ is redshifted,
since observer $\mathcal{\mathcal{O}}$ is moving away from the black-body
source. To obtain the blueshifted Doppler effect, one could flip the
sign on the velocity of $\mathcal{O}'$ with respect to the source
to become $-u$.

Since the Planck's law (\ref{eq:Plancklawnew-2}) contains the solid
angle term $d\Omega$, we need to know how it transforms under Lorentz
transformation. First, from (\ref{eq:doppler}) and (\ref{eq:doppler2-2})
we obtain:
\[
\gamma\left(1+\beta\cos\theta'\right)=\frac{1}{\gamma\left(1-\beta\cos\theta\right)}.
\]
Second, differentiating equation (\ref{eq:aberration-2}) or (\ref{eq:aberration-2-1})
with respect to a parameter will give:
\begin{equation}
\frac{d\theta'}{d\theta}=\frac{\omega}{\omega'}.\label{eq:doppler3}
\end{equation}
With (\ref{eq:doppler3}), the solid angle $d\Omega=\sin\theta d\theta d\phi$,
will transform as follows \citep{Johnson}:
\begin{equation}
\frac{d\Omega'}{d\Omega}=\left(\frac{\omega}{\omega'}\right)^{2},\label{eq:solidangle}
\end{equation}
using (\ref{eq:doppler0}), (\ref{eq:doppler3}), and the fact that
the azimuth angle are not affected by the transformation, $d\phi=d\phi'$.

\subsection{The Phase-Space Volume Transformation}

The next variable in the Planck's law that transformed under the Lorentz
transformation is $N$, the number or distribution of modes with frequency
$\omega$. To know how this variable changes under different inertial
frames, one needs to consider the phase-space volume transformation.
The phase-space volume element is not a Lorentz scalar, see a detailed
explanation in \citep{Debbasch}. First, let us obtain the transformation
for the 3-volume element of the spatial part of the phase-space. In
$\mathcal{O}$ and $\mathcal{O}'$, the infinitesimal 3-volume element
are defined as, respectively:
\begin{align}
d^{3}\mathbf{x} & =dx\wedge dy\wedge dz,\label{eq:volumeelement-2}\\
d^{3}\mathbf{x}' & =dx'\wedge dy'\wedge dz'.\nonumber 
\end{align}
To derive the transformation relation between these 2 elements, let
us consider a measurement of spatial length in $\mathcal{O}.$ To
measure spatial length, 2 events must be simultaneous in time with
respect to the observer. Let us consider 2 simultaneous events $\mathbf{p}=\left(t_{i},x_{i}\right)$
and $\mathbf{q}=\left(t_{f},x_{f}\right)$ along the $x$-axis at
time $t_{i}=t_{f}=0$, where $x_{i}=0$ and $x_{f}=\ell$. According
to $\mathcal{O},$ the length between these 2 event is:
\begin{align*}
\Delta x & =x_{f}-x_{i}=\ell.
\end{align*}
Now, let these 2 events be perceived in frame $\mathcal{O}'$, moving
with respect to $\mathcal{O}.$ Using Lorentz transformation (\ref{eq:Lorentz}):
\[
\begin{array}{cc}
t'_{i} & =\gamma\left(t_{i}-\beta x_{i}\right)=0,\\
x'_{i} & =\gamma\left(x_{i}-\beta t_{i}\right)=0,
\end{array}\;\begin{array}{cc}
t'_{f} & =\gamma\left(t_{f}-\beta x_{f}\right)=-\beta\gamma\ell,\\
x'_{f} & =\gamma\left(x_{f}-\beta t_{f}\right)=\gamma\ell;
\end{array}
\]
it is clear that in $\mathcal{O}'$ these 2 events are not simultaneous.
The spatial length between $\mathbf{p}$ and $\mathbf{q}$ in $\mathcal{O}'$
is: 
\begin{align*}
\Delta x' & =x'_{f}-x'_{i}=\gamma\ell.
\end{align*}

Now for our black-body case, we need to calculate how much modes are
inside a volume element. Let the finite volume element in $\mathcal{O}$
be $\Delta x\Delta y\Delta z$. This volume will be perceived simultaneously
by $\mathcal{O}$ at an instant time $t$, while in $\mathcal{O}',$
it will be perceived as the volume swept by the plane $\Delta y'\Delta z'=\Delta y\Delta z$
from $x'_{i}$ to $x'_{f}$ along a time interval $\Delta t'=t'_{f}-t'_{i};$
this is the physical interpretation of $\Delta x'\Delta y'\Delta z'.$
Taking the infinitesimal value $\Delta\mathbf{x}\rightarrow d\mathbf{x}$,
we have: 

\begin{equation}
d^{3}\mathbf{x}'=\gamma d^{3}\mathbf{x}.\label{eq:vol}
\end{equation}
Notice that the standard volume contraction formula is $d^{3}\mathbf{x}'=\gamma^{-1}d^{3}\mathbf{x},$
which has different interpretation with (\ref{eq:vol}). The standard
volume contraction is the comparison between volumes that are both
measured instantaneously with respect to each times on each frames,
while in (\ref{eq:vol}), the measurement of $d^{3}\mathbf{x}'$ is
done along a time interval $\Delta t'$.  For an alternative derivation
for this formula, one could consult \citep{Peebles,Heer,Henry}.

Second, let us obtain the transformation for the 3-volume element
in the momentum space, which are defined as follows:
\begin{align}
d^{3}\mathbf{p} & =dp_{x}\wedge dp_{y}\wedge dp_{z},\label{eq:volumeelement-1-1}\\
d^{3}\mathbf{p}' & =dp_{x}'\wedge dp_{y}'\wedge dp_{z}',\nonumber 
\end{align}
respectively for $\mathcal{O}$ and $\mathcal{O}'$. The 4-momentum
are constrained such that it's norm $\left|^{4}\mathbf{p}\right|=p_{\mu}p^{\mu}$
is constant:
\begin{equation}
p_{\mu}p^{\mu}=-p_{t}^{2}+p_{x}^{2}+p_{y}^{2}+p_{z}^{2}=-m^{2}c^{2}.\label{eq:energyconstraint}
\end{equation}
Differentiating this constraint with respect to any parameter gives:
\begin{equation}
dp_{t}=\frac{1}{p_{t}}\left(p_{x}dp_{x}+p_{y}dp_{y}+p_{z}dp_{z}\right).\label{eq:momentumconstraint-1}
\end{equation}
Let us consider the infinitesimal version of the transformation of
4-momentum; such transformations have similar forms with (\ref{eq:Lorentz-4}).
Inserting the infinitesimal version of (\ref{eq:Lorentz-4}) to (\ref{eq:volumeelement-1-1}),
and then using the constraint (\ref{eq:momentumconstraint-1}), gives:
\begin{align}
d^{3}\mathbf{p}' & =\gamma\left(1-\beta\frac{p_{x}}{p_{t}}\right)dp_{x}\wedge dp_{y}\wedge dp_{z}.\label{eq:hmm}
\end{align}
However, coefficient in (\ref{eq:hmm}) is simply $\frac{p_{t}'}{p_{t}}$
by (\ref{eq:Lorentz-4}), so one has:
\begin{align}
d^{3}\mathbf{p}' & =\frac{p_{t}'}{p_{t}}d^{3}\mathbf{p}=\frac{\omega'}{\omega}d^{3}\mathbf{p},\label{eq:mom}
\end{align}
by the photon momentum (\ref{eq:photon2}) and the Doppler effect
(\ref{eq:doppler3}). Now, using (\ref{eq:vol}) and (\ref{eq:mom}),
we can obtain the transformation of the phase-space volume element
between 2 inertial observers as follows:
\begin{align*}
d^{3}\mathbf{x}'\wedge d^{3}\mathbf{p}' & =\gamma\frac{\omega'}{\omega}d^{3}\mathbf{x}\wedge d^{3}\mathbf{p},\\
 & =\gamma^{2}\left(1-\beta\cos\theta\right)d^{3}\mathbf{x}\wedge d^{3}\mathbf{p}.
\end{align*}

\subsection{The Relativistic Distribution Function}

The trajectory of the photon in spacetime is described by a curve
in the Minkowski space. At an instantaneous (constant) time $t$,
one could define a hypersurface $\sigma_{t}$ and extend the hypersurface
to the phase-space by attaching the momentum space $T_{p}^{*}\sigma_{t}$
(the cotangent bundle of $\sigma_{t}.$) to $\sigma_{t}.$ The worldine
that cross the phase-space will mark a point on the phase-space $\sigma_{t}\times T_{p}^{*}\sigma_{t}$,
describing the state of the photon at time $t$. Moreover, one can
construct the phase-space volume element in the phase-space $\sigma_{t}\times T_{p}^{*}\sigma_{t}$,
namely $d^{3}\mathbf{x}\wedge d^{3}\mathbf{p}$. The relativistic
distribution function $f\left(\mathbf{x},\mathbf{p}\right)$ is defined
as the number of wordlines that cross the phase-space, i.e, the states,
per volume element \citep{Liboff}:
\[
f\left(\mathbf{x},\mathbf{p}\right)=\frac{dN}{d^{3}\mathbf{x}d^{3}\mathbf{p}}.
\]
Now, if we have another coordinate patch for the Minkowski space,
namely $\left(t',x',y',z'\right)$, it will perceive different time-constant
hypersurface, different phase-space, and different phase-space volume
element, namely $d^{3}\mathbf{x}'\wedge d^{3}\mathbf{p}'$. Since
the phase-space volume element changes, the number of wordlines that
cross the volume element will also change accordingly. However, the
number of worldlines (and the worldlines themselves) in $\mathcal{M}$
do not change by coordinate transformation. Hence, it is reasonable
to assume that the worldline density $f\left(\mathbf{x},\mathbf{p}\right)$
is invariant under coordinate (Lorentz) transformation, namely \citep{Liboff}:
\[
\frac{dN}{d^{3}\mathbf{x}d^{3}\mathbf{p}}=\frac{dN'}{d^{3}\mathbf{x}'d^{3}\mathbf{p}'}.
\]
With this reasonable assumption, one could have the number of states/modes
transformation between two inertial frames $\mathcal{O}$ and $\mathcal{O}'$,
which is:
\begin{equation}
\frac{dN'}{dN}=\frac{d^{3}\mathbf{x}'d^{3}\mathbf{p}'}{d^{3}\mathbf{x}d^{3}\mathbf{p}}=\gamma^{2}\left(1-\beta\cos\theta\right).\label{eq:numbertransf}
\end{equation}
Finally, with (\ref{eq:doppler2-2}), (\ref{eq:solidangle}), (\ref{eq:vol}),
and (\ref{eq:numbertransf}), we have all the ingredients to show
that the Planck's law are invariant under Lorentz transformation.

\subsection{Black-body Radiation in Moving Frame}

The objective in this section is to know if the Planckian distribution
is invariant under Lorentz transformation. To avoid the problem of
temperature in moving bodies, we use the Planck's Law in the form
of equation (\ref{eq:Plancklawnew-2}), that is, the one with the
term containing $\omega{}_{\mathrm{max}}'$, instead of the original
form (\ref{eq:Plancklaw5}) with the term containing $\kappa_{B}T$.
The derivation in this section is based on the derivation in \citep{Henry},
with some slight modifications. Inserting the relativistic Doppler
effect (\ref{eq:doppler2-2}), the transformation of solid angle (\ref{eq:solidangle}),
the volume contraction (\ref{eq:vol}), and the transformation of
the number of modes distribution (\ref{eq:numbertransf}) to (\ref{eq:Plancklawnew-2}),
we obtain: \begin{widetext}
\[
\frac{dN'}{\gamma^{2}\left(1-\beta\cos\theta\right)}=\frac{\left(\omega'\right)^{2}}{4\pi^{3}c^{3}}\frac{1}{\gamma^{2}\left(1-\beta\cos\theta\right)^{2}}\frac{1}{\exp\left(w\frac{\omega'}{\gamma\left(1-\beta\cos\theta\right)}\frac{\gamma\left(1-\beta\cos\theta\right)}{\omega{}_{\mathrm{max}}'}\right)-1}\frac{d\omega'}{\gamma\left(1-\beta\cos\theta\right)}\left(\frac{\omega'}{\omega}\right)^{2}d\Omega'\frac{1}{\gamma}d^{3}\mathbf{x}',
\]
 \end{widetext} that can be simplified as:
\begin{equation}
dN'=\frac{\left(\omega'\right)^{2}}{4\pi^{3}c^{3}}\frac{1}{\exp\left(w\frac{\omega'}{\omega'_{\mathrm{max}}}\right)-1}d\omega'd\Omega'dV',\label{eq:Plancklawnewmoving}
\end{equation}
which is exactly the form of Planck's law (\ref{eq:Plancklawnew-2}).
All the variables inside the equation transform according to the Lorentz
transformation, but the relation between these variables is invariant,
hence, the observer in a moving frame will still observe the Planckian
spectrum. However, there is a subtlety in the relation (\ref{eq:Plancklawnewmoving}).
$\omega{}_{\mathrm{max}}'$, the 'maximal' frequency, also transformed
as the angular frequency via the relativistic Doppler effect (\ref{eq:doppler2-2}).
However , $\omega{}_{\mathrm{max}}'$ is the 'maximal' frequency in
the distribution (\ref{eq:Plancklawnew-2}) of a black-body in the
rest frame, observed by a moving observer. It is not necessarily the
'maximal' frequency obtained from the distribution (\ref{eq:Plancklawnewmoving})
of a black-body observed by the moving observer. We will see that
they are equivalent as follows. Let us multiply the LHS and RHS of
(\ref{eq:Plancklawnewmoving}) with $\hbar\omega'$ to obtain:
\begin{equation}
dE'=\underset{\rho'_{E}\left[\omega'\right]}{\underbrace{\frac{\hbar\left(\omega'\right)^{3}}{4\pi^{3}c^{3}}\frac{1}{\exp\left(w\frac{\omega'}{\omega{}_{\mathrm{max}}'}\right)-1}}}d\omega'd\Omega'dV'.\label{eq:Plancklawnewmoving-1-1}
\end{equation}
$E'$ is the energy of modes with frequency $\omega'$ observed in
the moving frame $\mathcal{O}'$. Let us find the value of $\omega'$
that maximize $E'$ from the equation $\frac{d\rho'_{E}\left[\omega'\right]}{d\omega'}=0$:
\begin{widetext}
\begin{equation}
\frac{\hbar\left(\omega'\right)^{2}}{4\pi^{3}c^{3}}\frac{1}{\exp\left(w\frac{\omega'}{\omega{}_{\mathrm{max}}'}\right)-1}\left(3-w\frac{\omega'}{\omega{}_{\mathrm{max}}'}\frac{\exp\left(w\frac{\omega'}{\omega{}_{\mathrm{max}}'}\right)}{\exp\left(w\frac{\omega'}{\omega{}_{\mathrm{max}}'}\right)-1}\right)=0.\label{eq:dplanckmov}
\end{equation}
 \end{widetext} Defining $w'=w\frac{\omega'}{\omega{}_{\mathrm{max}}'}$
as in the previous subsection, the equation from (\ref{eq:dplanckmov})
that we need to solve is then:
\begin{equation}
w'=3\left(1-e^{-w'}\right),\label{eq:Lambertmov-1}
\end{equation}
which is exactly similar to (\ref{eq:Lambert}). Hence it posses a
same solution:
\[
w'=3+W_{0}\left(-3e^{-3}\right),
\]
namely, the Wien coefficient is invariant under Lorentz transformation:
$w'=w$. Then, if $\omega'=\omega'_{\mathrm{max}}$ is the solution
to (\ref{eq:dplanckmov}), we have: 
\[
\frac{w'}{w}=\frac{\omega'_{\mathrm{max}}}{\omega{}_{\mathrm{max}}'}=1,
\]
or namely $\omega'_{\mathrm{max}}=\omega{}_{\mathrm{max}}'$: the
Lorentz transformation sends 'maximal' frequency in one inertial frame,
to a  'maximal' frequency in another inertial frame. (\ref{eq:Plancklawnewmoving-1-1})
becomes:
\begin{equation}
dN'=\frac{\left(\omega'\right)^{2}}{4\pi^{3}c^{3}}\frac{1}{\exp\left(w\frac{\omega'}{\omega{}_{\mathrm{max}}'}\right)-1}d\omega'd\Omega'dV'.\label{eq:Plancklawnewmoving-2}
\end{equation}
With this, we can state that the Planck's radiation law in the form
of (\ref{eq:Plancklawnew-2}) is invariant under Lorentz transformation.
This result is similar to \citep{Henry}.

\end{document}